\documentstyle[eqsecnum,aps,prb,multicol,epsf]{revtex}
\begin{document}
\def\pmb#1{\setbox0=\hbox{#1}%
  \kern-.025em\copy0\kern-\wd0
  \kern.05em\copy0\kern-\wd0
  \kern-.025em\raise.0433em\box0 }
\newcommand{\rvec}{{\bf r}}
\newcommand{\zhat}{\hat{\bf z}}
\newcommand{\bftau}{\pmb{$\tau$}}

\title{Interstitials, Vacancies and Dislocations in Flux-Line Lattices:
A Theory of Vortex Crystals, Supersolids and Liquids}

\author{M. Cristina Marchetti}
\address{Physics Department, Syracuse University, Syracuse, NY 13244}

\author{Leo Radzihovsky}
\address{Physics Department, University of Colorado, Boulder, CO 80309}

\date{\today}

\maketitle

\begin{abstract} 
We study a three dimensional Abrikosov vortex lattice in the presence
of an equilibrium concentration of vacancy, interstitial and
dislocation loops.  Vacancies and interstitials renormalize the
long-wavelength bulk and tilt elastic moduli. Dislocation loops lead
to the vanishing of the long-wavelength shear modulus.  The coupling
to vacancies and interstitials - which are always present in the
liquid state - allows dislocations to relax stresses by climbing out
of their glide plane. Surprisingly, this mechanism does not yield any
further independent renormalization of the tilt and compressional
moduli at long wavelengths. The long wavelength properties of the
resulting state are formally identical to that of the ``flux-line
hexatic'' that is a candidate ``normal'' hexatically ordered vortex
liquid state.

\end{abstract}

\begin{multicols}{2}

\section{Introduction}
\label{introduction}
\vspace{-0.1cm}
Both disorder and thermal fluctuations strongly affect the properties
of the vortex array induced in type II superconductors by an external
magnetic field.\cite{huse_radzihovsky,blatter,brandt_review}. One of
the most striking consequences of thermal fluctuations, particularly
pronounced in the high-$T_c$ materials, is the resistive vortex liquid
state,\cite{nelson_seung} located between the $H_{c2}(T)$ line and the
vortex solid in the magnetic field (H)-temperature (T) phase
diagram.\cite{Hc1comment} Upon field cooling, a vortex liquid freezes
into an Abrikosov vortex solid. The nature of the freezing transition
and of the resulting vortex solid phase depends on the amount of
disorder present in the material.  In dirty samples the vortex solid
has been described as a ``vortex glass'',\cite{FFH} and its
translational correlation length is limited by disorder to a {\em
finite} value.\cite{Larkin} In three dimensions, the low temperature
vortex glass solid is expected to be a true superconductor with a
vanishing {\em linear} resistivity. For weak disorder, the vortex
solid state is expected to be a topologically ordered ``Bragg'' glass
state in three dimensions, with logarithmically growing vortex
displacements, but bound dislocation loops.\cite{BraggGlass} The
freezing transition of the vortex array has been observed to be first
order\cite{Gammel,Zeldov} in ultra clean samples and continuous in
dirty
superconductors.\cite{Gammel,Koch,Zeldov,melting,continuous_melting}

In very clean samples, where the disorder-limited translational
correlation length is thousands of intervortex lattice constants, the
low temperature phase can be well approximated by a vortex {\em
lattice}. Within an elastic description, the
primary low temperature excitations of the vortex lattice are phonons,
characterizing two-dimensional displacements of vortex lines from their
preferred lattice positions. As the temperature (or field) is raised
towards the melting transition, other excitations
become important. By definition, these are {\em defects} in the vortex
lattice, i.e. they are not describable in terms of single-valued
vortex displacements.  These line defects are dislocations,
disclinations, vacancies and interstitials and must be included in
the model for a complete description of the melting of the vortex solid
and of the properties of the resulting vortex liquid state.

At low temperatures, in a well-ordered Abrikosov lattice
state,\cite{BraggGlass} these defects are bound, as the energy of an
isolated {\em line} defect diverges with system size. At higher
temperatures entropy can, however, drive a proliferation of these line
defects, in analogy with the melting of two-dimensional
solids.\cite{KT,dsfisher} Two melting scenarios are
possible:\cite{melting} (i) a one-stage first order transition from a
solid to an {\em isotropic} vortex liquid where {\em both}
dislocations and disclinations unbind simultaneously (as it occurs in
the melting of ordinary 3d solids), (ii) a two-stage, possibly
continuous transition\cite{continuous_melting}, where dislocations
unbind first, leading to a hexatic flux-line liquid with residual
bond-orientational order, vanishing shear modulus, but finite hexatic
stiffness. This first transition would then be followed by a
proliferation of disclination loops, thereby completing the transition
into an {\em isotropic} vortex liquid. The first stage of this second
scenario for the melting of the Abrikosov lattice was first suggested
by Marchetti and Nelson.\cite{mcmdrn} While avoiding the subtle
question of the melting transition itself, they adapted the method
developed long ago by Nelson and Toner\cite{NelsonToner} to describe
the vortex line-liquid state. Marchetti and Nelson described the
hexatic vortex liquid as a vortex lattice with an equilibrium
concentration of dislocation loops, treating the latter in the
Debye-Huckle approximation. Through detailed calculations, they
demonstrated that dislocations drive the long wavelength shear modulus
of the system to zero and computed the effective hexatic stiffness of
the resulting orientationally ordered vortex liquid.\cite{mcmdrn}

Vacancies and interstitials constitute another class of defects that play an
important role in solids. In ordinary crystals and in 2d vortex
lattices, these {\em point} defects cost finite energy and are
therefore present in finite density, at {\em any finite}
temperature. While their static effects in these systems are minimal,
their density represents an important hydrodynamic mode which must be
included in the correct description of crystal
hydrodynamics.\cite{martin}

In strong contrast, in vortex lattices, vacancies and interstitials
are {\em line} defects with energy proportional to their length and
thereby diverging in bulk samples. Hence one expects that at low
temperatures their average density vanishes. At higher temperatures, this
positive energetic contribution to the free energy can, however,  
be compensated
by a negative entropic contribution associated with line wandering, which
also scales with the defect length, in analogy with
the Kosterlitz-Thouless picture.\cite{KT}  These considerations allow two
thermodynamically distinct crystalline phases, with a sharp phase
transition between them. While in both phases dislocations and
disclinations are bound, and as a consequence there is long-range
translational order (true Bragg spots in an X-ray scattering experiment)
and a finite shear modulus, 
line vacancies and interstitials are bound in the low temperature crystal, 
but have proliferated in the high temperature crystal phase. A
thermodynamically sharp distinction between these two crystal phases
in three-dimensional vortex systems was first emphasized by
M.P.A. Fisher and Lee\cite{MPAFisherLee} based on the mathematical
correspondence between vortex lines and world-lines of two-dimensional
quantum bosons. In this mapping the low temperature vortex crystal
maps onto a 2d Wigner crystal and the high temperature vortex solid
corresponds to the quantum supersolid phase of 2d bosons, with
vacancies and interstitials in its ground state. The quantum
supersolid is quite exotic, in that it is both crystalline and
superfluid. Correspondingly, due to the finite density of vacancies
and interstitials in the vortex supersolid, vortex lines can move
arbitrarily far and entangle, as in a vortex liquid, and therefore
this phase exhibits finite linear resistivity.

While experiments seem to rule out the existence of an equilibrated
vortex supersolid phase in bulk 2d {\em quantum} crystals, based on
detailed calculations, Frey, Nelson and D.S. Fisher,\cite{fnf} have
argued that such a phase is more likely to exist in flux-line
arrays at high fields because of the layered structure of high
temperature superconductors.  These authors conclude that a vortex
supersolid phase will certainly exist in anisotropic superconductors
for magnetic fields above a decoupling field $B_x$ where vortices in
different CuO$_2$ layers are essentially decoupled by thermal
fluctuations.\cite{glazman} Furthermore, even if an equilibrium
supersolid phase was absent, an appreciable nonequilibrium density of
vacancies and interstitials may still be present in a flux-line
lattice, when the vortex array undergoes a first order freezing
transition upon cooling in a constant field.

In this paper we study the effects of vacancies and interstitials
within the vortex supersolid and liquid phases. As discussed above, in
the supersolid phase, aside from being responsible for its existence,
these defects provide a mechanism for vortex line wandering and
consequently for its finite resistivity. They also are important
degrees of freedom, in addition to phonons, that must be incorporated
in the correct description of the supersolid. In Sec.\ref{supersolid}
we construct a model of a vortex supersolid, as an elastic lattice
with an equilibrium concentration of unbound vacancies and
interstitials. We compute the flux-line density correlation functions,
that characterize the static equilibrium properties of this phase, and
extract from these the effective elastic moduli of the supersolid
phase. We find that the long-wavelength shear modulus is unaffected by
the {\em fluctuations} in the density of vacancies and
interstitials. This result is consistent with the
vacancy-interstitials' inability to relax shear, and confirms the
finiteness of the supersolid shear rigidity which distinguishes it
from a vortex liquid.  We also compute a finite, downward
renormalization of the compressional and tilt moduli by vacancy and
interstitial density fluctuations. We demonstrate that as a
consequence of the reduction of the effective tilt modulus, the
flux-line {\em wandering} is enhanced and, analogously to a vortex
liquid, a bulk vortex supersolid is always entangled, i.e., it does
not exhibit longitudinal superconductivity.

The existence of a vortex supersolid allows for two scenarios for the
melting transition into the vortex liquid state. At low magnetic
fields, we expect\cite{fnf} this transition to be directly from the
low temperature, non-supersolid crystal into a vortex liquid phase.
As discussed above, the vortex liquid state itself can be either a
fully disordered isotropic liquid or a bond-orientationally ordered
liquid that can further disorder into an isotropic liquid via
disclination unbinding. %
\begin{figure}[bth]
{\centering
\setlength{\unitlength}{1mm}
\begin{picture}(150,63)(0,0)
\put(-34,-73){\begin{picture}(150,63)(0,0) 
\includegraphics{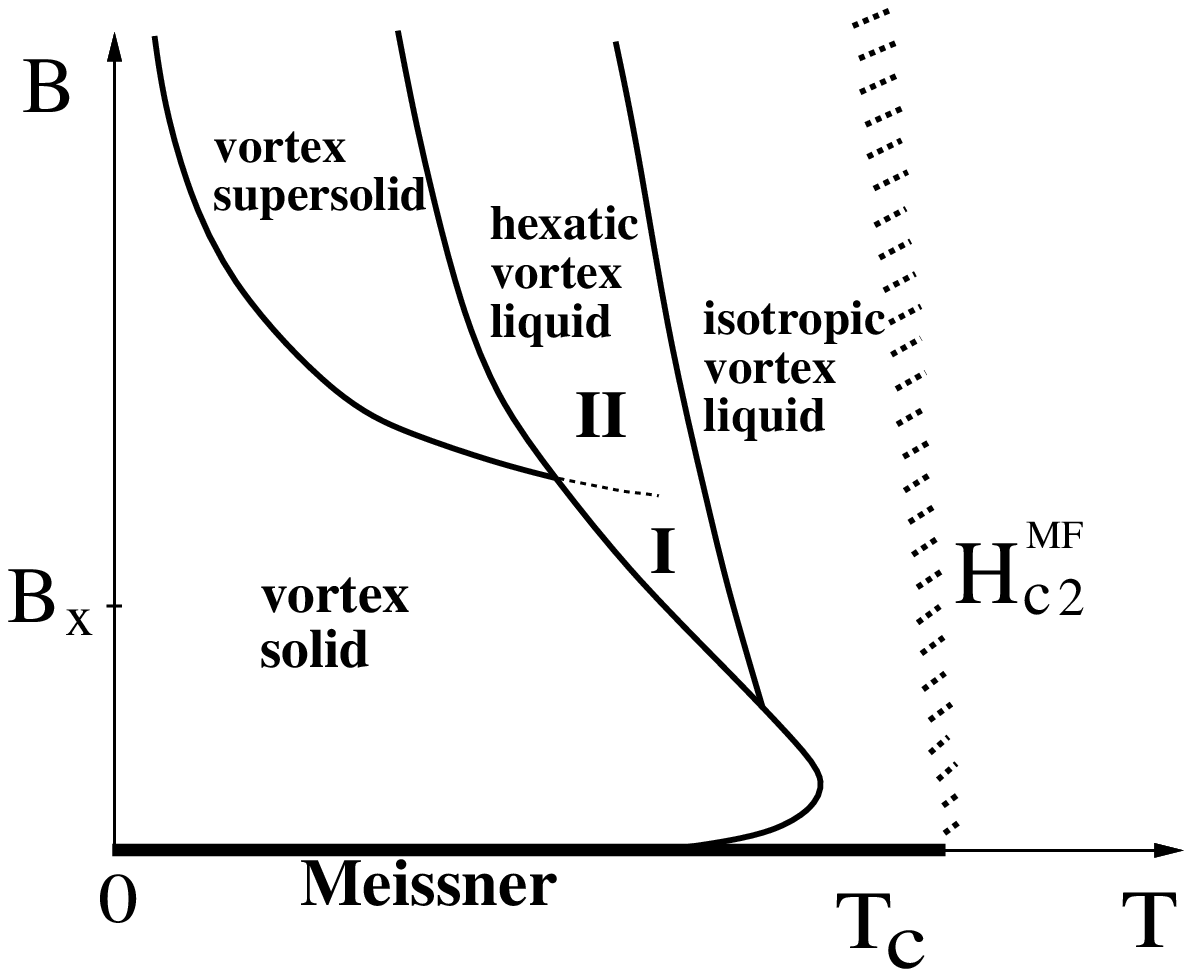}
\end{picture}}
\end{picture}}
Fig.1.{Schematic $B-T$ phase diagram illustrating the flux line
crystal, vortex ``supersolid'' (guaranteed to exist at fields much
higher than the decoupling field $B_x$), the orientationally ordered
hexatic vortex liquid, and the isotropic vortex liquid.}
\end{figure}
Alternatively, the non-supersolid crystal can first undergo a
transition into a vortex supersolid by a proliferation of vacancies
and interstitials,\cite{fnf} and subsequently melt into a vortex
liquid.\cite{fb} These two scenarios have been dubbed as type I and
type II melting, respectively,\cite{fnf} and are illustrated in Fig.1.

Very recent experiments investigating vortex penetration through
surface barriers in the presence of a transport current in clean BSCCO
samples have indeed indicated that the nature of the melting line may
change at high field, even in the absence of point
disorder.\cite{fuchs} In fact these experiments suggest the existence
of an intermediate phase between the solid and the liquid at high
fields.  This observed regime lies above the first order melting line
determined by equilibrium magnetization measurements, but
apparently exhibits a certain amount of sixfold periodicity, as
indicated by neutron scattering.\cite{forgan} It is therefore a
candidate for either the supersolid or the hexatic liquid phases
studied here.

Although the nature of the type I and type II melting transitions
should be quite different,\cite{melting_comment} in either case
vacancies and interstitials should proliferate in the resulting vortex
liquid state.\cite{defects_comment} Hence, in addition to
dislocations, vacancy and interstitial defects {\em must} be taken
into account for the proper description of a vortex liquid.  They were
not, however, explicitly included in the calculations of
Ref.\onlinecite{mcmdrn}.  When vacancies and interstitials are absent,
dislocation loops in the vortex line-lattice are restricted to lie in
the plane defined by the magnetic field axis and their Burger's
vector, and can only glide (see Fig.5), but not climb. One of the
consequences of this is that the effective tilt modulus of the vortex
liquid is not renormalized by dislocations and in the model of
Ref. \onlinecite{mcmdrn} is therefore identical to that of the vortex
lattice.

The results presented in Sec.\ref{liquid} remedy the limitations of
Ref.\onlinecite{mcmdrn}, by explicitly including vacancies and
interstitials in the description of a vortex liquid.  Vacancies and
interstitials renormalize the tilt and compressional moduli and allow
dislocation loops to climb out of their glide plane by absorbing and
emitting these defects.  At long wavelengths, however, the coupling of
dislocations to vacancies and interstitials does not yield any {\em
additional independent} renormalization of the tilt and compressional
moduli. Whether this is a general property of the vortex lattice, or
an artifact of the quadratic model and the Debye-Huckel approximation
used here, remains an open question. Our work yields a complete
description of a bond-orientationally ordered hexatic vortex liquid,
whose tilt and bulk moduli are renormalized by dislocations, vacancies
and interstitials, and whose shear modulus is driven to zero by the
proliferation of these defects.

Our results characterizing the properties of the various phases shown
in Fig.1 are summarized in the two tables below.  We stress that
hexatic I and hexatic II are not two distinct hexatic phases, but
rather two different regimes within the same hexatic phase,
distinguished by a high and low concentration of vacancy/interstitial
defects, respectively. The crossover between these two regimes is
indicated in Fig.1 by a dotted line. The results quoted in Table 1 for
the hexatic I phase are those obtained by Marchetti and Nelson in
Ref.\onlinecite{mcmdrn} assuming that no vacancy/interstitial defects
are present in the system. These results, however, only apply well
below the dotted line in Fig.1, very close to the solid-hexatic phase
boundary. As the transition to the isotropic liquid is approached, the
large number of interacting dislocation loops present in the hexatic
will invitably generate vacancy/interstitial defects as well, leading
to the breakdown of the model of Ref.\onlinecite{mcmdrn}.  The type of
long-range order present or absent in each of these phases is
summarized qualitatively in Table 2.

\end{multicols}
\begin{center}
{\begin{tabular}{|c||c|c|c|c|}\hline
& & & & \\
&~shear~ &~compression~ & tilt & $\langle W^2\rangle$ (Eq. (3.28))\\ 
& & & & \\\hline\hline
& & & & \\
crystal & $c_{66}$ & $c_{11}$ & $c_{44}$ & $0$\\ 
& & & & \\\hline
& & & & \\
~supersolid~ & $c_{66}$ & ~${c_{11}\chi^{-1}-\gamma^2\over c_{11}+\chi^{-1}+2\gamma}$~ &
   ~${c_{44}K-\gamma'^2\over c_{44}+K+2\gamma'}$~ & ${n_0^2k_BT\over K}$\\ 
& & & & \\\hline
& & & & \\
hex II & $0$ & $c_{11}-c_{66}$ & ${c_{44}K-\gamma'^2\over c_{44}+K+2\gamma'}$  & ~${n_0^2k_BT\over c_{44}}\Big[1+{(c_{44}-\gamma')^2\over c_{44}K-\gamma'^2}\Big]$~ \\ 
& & & & \\\hline
& & & & \\
hex I & $0$ & $c_{11}-c_{66}$ & $c_{44}$  & ${n_0^2k_BT\over c_{44}}$\\ 
& & & & \\\hline
\end{tabular}}
\end{center}

\vspace{0.1in}
Table 1. A summary of our results for the elastic constants and the winding number in the various
phases of Fig.~1.
\vspace{0.1in}

\begin{center}
{\begin{tabular}{|c||c|c|c|}\hline
&Translational & Orientational & Longitudinal \\ 
& LRO & LRO & Superconductivity \\ \hline\hline
crystal & yes & yes & yes \\\hline
supersolid & yes & yes & no$^*$\\\hline
hex I & no & yes & no$$\\\hline
hex II & no & yes & no\\\hline
liquid & no & no & no\\\hline
\end{tabular}}
\end{center}

\vspace{0.1in}
Table 2. This table displays the presence or absence of translational
and orientational long-range order (LRO) as well as longitudinal
superconductivity in each phase. The asterisk serves to emphasize that
although the supersolid does not exhibit longitudinal
superconductivity, the degree of screening of longitudinal currents in
this phase is substantially different from that of the vortex liquid
phase (see Table 1).

\begin{multicols}{2}

\section{Elastic properties of defect-free vortex lattices}
\label{crystal}
\subsection{Model}
\label{model_lattice}

We begin by recalling the properties of an ordered, {\em defect-free}
vortex lattice, which we expect to be stable at low temperature and
field in a clean superconductor.\cite{BraggGlass} As discussed in the
Introduction, long-scale degrees of freedom of this system are
uniquely characterized by a single-valued vortex displacement field,
${\bf u}({\bf r})$. With the convection in which the external magnetic
field ${\bf H}_0$ (aligned with the $c$ axis of the superconductor)
points along the $z$ axis, the two-dimensional vector displacement
${\bf u}({\bf r}_\perp,z)$ is confined to the $xy$ plane.

The long-wavelength properties of a triangular flux-line lattice are
characterized by the elastic free energy functional
\begin{equation}
\label{freeel}
F_{\rm latt}={1\over 2}\int d{\bf r}\Big
[2c_{66}u_{ij}^2+(c_{11}-2c_{66})u_{kk}^2 +c_{44}(\partial_z{\bf
u})^2\Big],
\end{equation}
where 
\begin{equation}
\label{symstrain}
u_{ij}={1\over 2}(w_{ij}+w_{ji})
\end{equation}
is the symmetrized two-dimensional strain tensor, with
\begin{equation}
\label{strain}
w_{\alpha j}=\partial_\alpha u_j,
\end{equation}
a $3\times2$ hybrid strain tensor. Greek indices take on the
full three-dimensional set of labels $x,y,z$, and Latin indices are
reserved for the purely two-dimensional set $x,y$. We will use this
notation throughout the manuscript. The parameters $c_{66}$, $c_{11}$
and $c_{44}$ appearing in the Eq.\ref{freeel} are the shear,
compressional and tilt modulus, respectively.\cite{notation_comment}
In contrast to ordinary crystals, in a flux-line lattice vortex
interactions extend over a range of order of the London penetration depth,
$\lambda$, which can be quite large, especially in high T$_c$
superconductors. As emphasized in the extensive literature on the
subject,\cite{brandt,blatter} on scales shorter than $\lambda$, this
leads to wavevector-dependent elastic moduli. For a detailed
comparison with experiments, inclusion of these nonlocal elastic
effects can be important, especially at high fields, and they can be
easily incorporated into our results by replacing all of the {\em
bare} elastic moduli by the proper wavevector-dependent
expressions.\cite{dsfisher_elasticity}

When the lattice contains no vacancies nor interstitials, the number of
flux lines equals the number $N$ of sites in the triangular
lattice.  On the average, the flux lines are aligned with
the external field and the equilibrium magnetic flux density field is
given by ${\bf B}_0=\hat{\bf z}B_0=\hat{\bf z}\phi_0 n_0$, where
$\phi_0=hc/2e$ is the flux quantum, $n_0=N/A\equiv1/a_c$ is the
equilibrium number density of vortex lines and $a_c$ the area of the
primitive unit cell.

Fluctuations in the local induction $\delta {\bf B}({\bf r})= {\bf
B}({\bf r})-{\bf B}_0$ can be described in terms of fluctuations in
the areal density of flux lines, $\delta n({\bf r}_\perp,z)=n({\bf
r}_\perp,z)-n_0$, and of a tilt vector field ${\bf t}({\bf
r}_\perp,z)$, with the relation 
\begin{eqnarray}
\label{Bz}
& &\delta B_z=\phi_0\delta n, \\
\label{Bperp}
& &{\bf B}_\perp=\phi_0{\bf t},
\end{eqnarray}
valid in the long wavelength $q\lambda<<1$ limit.\cite{foota}  In the
absence of vacancies and interstitials, the areal density of flux
lines and their orientation relative to the applied field direction
are entirely determined by the local strains, according to
\begin{eqnarray}
\label{densityfl}
& &\delta n/n_0=-\delta A/A=-u_{ii},\\
\label{tiltfl}
& &{\bf t}/n_0=\partial_z{\bf u}.
\end{eqnarray}
The condition $\bbox{\nabla}\cdot{\bf B}=0$ translates into a ``continuity''
constraint for the flux lines,
\begin{equation}
\label{continuity}
\partial_z\delta n+\bbox{\nabla}_\perp\cdot{\bf t}=0.
\end{equation}
As can be seen from Eqs.\ref{densityfl} and \ref{tiltfl}, 
this continuity constraint is identically
satisfied in the {\em
defect-free} vortex lattice, where the displacement $\bf u$
is single-valued.

\subsection{Correlation and response functions}
\label{CRcrystal_functions}

Thermal fluctuations in the density and tilt field are characterized
by the density-density correlation function (the structure factor),
\begin{equation}
\label{Sq}
S(q_\perp,q_z)={1\over V}\langle\delta n({\bf q})\delta n(-{\bf
q})\rangle,
\end{equation}
and the tilt field correlation function,
\begin{eqnarray}
\label{Tq}
T_{ij}(q_\perp,q_z)&=&{1\over V}\langle t_i({\bf q}) t_j(-{\bf q})\rangle\\\nonumber
&=& T_L(q_\perp,q_z)P^L_{ij}({\bf \hat{q}_\perp})
   +T_T(q_\perp,q_z)P^T_{ij}({\bf \hat{q}_\perp}),
\end{eqnarray}
where 
\begin{eqnarray}
P^L_{ij}({\bf \hat{q}_\perp})&=&\hat{q}_{\perp i}\hat{q}_{\perp j}\;,
\label{PL}\\
P^T_{ij}({\bf \hat{q}_\perp})&=&\delta_{ij}-
\hat{q}_{\perp i}\hat{q}_{\perp j}\;,
\label{PT}
\end{eqnarray}
are the longitudinal and the transverse projection operators,
respectively, ${\bf \hat{q}}_\perp={\bf q}_\perp/q_\perp$, and $V$ is
the volume of the superconductor.  In light of the constraint,
Eq.\ref{continuity}, the longitudinal part of the tangent field
correlator is proportional to the structure function, with
\begin{equation}
\label{tl}
T_L(q_\perp,q_z)={q_z^2\over q^2_\perp}S(q_\perp,q_z).
\end{equation}

The brackets $\langle\ldots\rangle$ in above expressions and
throughout the paper indicate a thermal average with a Boltzmann
weight $e^{-F/k_BT}/Z$, with $F$ the free energy functional and
$Z=\mbox{Trace}[e^{-F/k_BT}]$ the corresponding partition function. In
the defect-free (non-supersolid) vortex lattice, $F$ is given by
$F_{\rm latt}$, Eq.\ref{freeel}, and we denote the corresponding thermal
averages by $\langle\ldots\rangle_0$. Using
Eqs.\ref{densityfl} and \ref{tiltfl}, the structure function and the tilt field
correlation function can be expressed in terms of thermal averages of
the phonon field $\bf u$, and are therefore easily computed, 
with the result,
\begin{eqnarray}
\label{Sq0}
S^0({\bf q})& =&{q_\perp^2 n_0^2\over V}\langle|\hat{\bf q}_\perp\cdot{\bf u}({\bf q})|^2\rangle_0\nonumber\\
&=& {n_0^2k_BTq_\perp^2\over c_{11}q_\perp^2+c_{44}q_z^2},
\end{eqnarray}
and 
\begin{eqnarray}
\label{Tq0}
T^0_T({\bf q})& =&{q_z^2 n_0^2\over V}P^T_{ij}({\bf \hat{q}_\perp})\langle u_i({\bf q})u_j(-{\bf q})\rangle_0\\\nonumber
&=& {n_0^2k_BTq_z^2\over c_{66}q_\perp^2+c_{44}q_z^2}.
\end{eqnarray}

The structure function $S({\bf q})$ can be probed in a
neutron scattering experiment. The tilt correlation function
$T_{ij}({\bf q})$ is directly connected to the experimentally measurable
linear magnetic susceptibility tensor, $\chi_{ij}({\bf q})$,
according to
\begin{equation}
\label{chiij}
4\pi\chi_{ij}(q_\perp,q_z)=
-\delta_{ij}
+{\phi^2_0\over 4\pi k_BT}T_{ij}(q_\perp,q_z).
\end{equation}
Equation \ref{chiij} holds in the long wavelength $q\lambda <<1$
limit.  A more general relationship between susceptibility and tilt
correlation function that applies at scales shorter than $\lambda$ in
an anisotropic material can be found for instance in
Ref. \onlinecite{pbmcm}.  The first term on the right hand side of Eq.\ref{chiij} represents a
perfect diamagnetic (negative) Meissner response, which, in a mixed
state is considerably reduced by the ``normal'' paramagnetic vortex
tilt response $T_{ij}$, contained in the second term.
The linear susceptibility relates the
transverse flux density $\delta{\bf B}_{\perp}({\bf q})$ induced by an
external perturbation field $\delta{\bf H}_\perp({\bf q})$, applied
perpendicular to the field ${\bf H}_0=\hat{\bf z}H_0$ responsible for
the onset of the vortex state, with
\begin{equation}
\delta B_{\perp i}({\bf q})=[\delta_{ij}+4\pi\chi_{ij}({\bf q})]
\delta H_{\perp j}({\bf q})\;.
\end{equation}

From Eq.\ref{Tq0}, appropriate for a perfect, defect-free flux-line
lattice, we observe that the long wavelength limit of the transverse
part of the tilt field correlation function is {\em nonanalytic}, with
\begin{eqnarray}
\label{tiltlong0}
& & T^0_T(q_\perp=0,q_z)={n_0^2k_BT\over c_{44}}\;,\\
\label{rhos}
& & T^0_T(q_\perp,q_z=0)=0\;.
\end{eqnarray}
The nonanalyticity of the tilt correlation function reflects a
drastically different linear response of the defect-free vortex
lattice to two types of transverse field
perturbations.\cite{chenteitel}

The transverse field response in the limit $q_\perp<<q_z\rightarrow
0$, corresponds to a tilt perturbation of the flux lines, induced by
an applied transverse field $V\delta_{{\bf q}_\perp,{\bf 0}}\delta{\bf
H}_\perp(q_z)$ that is spatially homogeneous in the $xy$ plane.  The
corresponding long wavelength transverse susceptibility,
$\chi_T=P^T_{ij}\chi_{ij}$, is given by
\begin{equation}
\label{transverseM}
\lim_{q_z\rightarrow 0}\chi_T^0(q_\perp=0,q_z)=-{1\over 4\pi}
\Big[1-{B^2\over 4\pi c_{44}}\Big].
\end{equation}
If the second term in brackets on the right hand side of
Eq.\ref{transverseM} were absent, the superconductor would exhibit
perfect screening of the transverse perturbation.  Such a behavior can
for instance occur in flux-line arrays pinned by aligned damage
tracks. In the presence of such correlated disorder, the vortex
lattice is replaced by a thermodynamically distinct ``Bose'' glass
phase,\cite{BoseGlass} that is characterized by a divergent tilt
modulus $c_{44}$ and exhibits a transverse Meissner effect, with
$\lim_{q_z\rightarrow 0}\chi_T(q_z)=-1/4\pi$.  In the absence of {\em
anisotropic} pinning of the vortex lattice, such a perfect transverse
diamagnetic response is spoiled by the finite vortex tilt response,
that leads to only a partial screening of the transverse field,
displayed in Eq.\ref{transverseM}.

Conversely, the limit $q_z<<q_\perp\rightarrow 0$, describes a
magnetic response to a transverse field $V\delta_{q_z,0}\delta{\bf
H}_\perp({\bf q}_\perp)$ that is homogeneous along the z-axis, but is
spatially varying in the $xy$-plane. The induced $z$-directed screening
currents lead to a shear perturbation of the flux-line array, with the
response in a defect-free lattice given by
\begin{equation}
\lim_{q_\perp\rightarrow 0}\chi_T^0(q_\perp,q_z=0)=-{1\over 4\pi}.
\end{equation}
Thus the flux lattice exhibits perfect screening in response to this
$z$-independent transverse perturbation. Since the screening currents
involved in the shear perturbation run parallel to the applied field
${\bf H}_0\parallel\hat{\bf z}$, a perfect Meissner response to a
shear perturbation has also been termed longitudinal
superconductivity. It follows directly from the fact that, in contrast
to a liquid, a vortex lattice is characterized by a finite shear
modulus $c_{66}$.

For comparison, we recall that in a flux-line liquid, the long
wavelength limit of the transverse part of the tilt correlation
function is analytic, with
\begin{eqnarray}
\lim_{q_\perp\rightarrow 0}T_T^{\rm liquid}(q_\perp,q_z=0)&=&
\lim_{q_z\rightarrow 0}T^{\rm liquid}_T(q_\perp=0,q_z)\;,\nonumber\\
&=&{n_0^2k_BT\over c_{44}^{\rm liquid}}\label{tiltliquid}\;.
\end{eqnarray}
As expected, a vortex liquid, being {\em qualitatively} identical to
the normal state, albeit highly conductive, exhibits neither
transverse Meissner effect, nor longitudinal superconductivity. We
will return to this point again in Sec.\ref{liquid}.

Finally, we note that in the latter $q_z<<q_\perp\rightarrow 0$ limit,
the transverse part of the tilt correlation function corresponds to
the world-lines winding number $\langle W^2\rangle$ studied by Pollock
and Ceperley\cite{pollock} in their path integral approach to the
superfluid transition in quantum boson systems. In such an approach,
dating back to Feynman, the superfluid phase is identified with an
entangled state of boson world-line trajectories,
\begin{eqnarray}
\label{winding}
\lim_{q_\perp\rightarrow 0}T_T(q_\perp,q_z=0)&=&\langle
W^2\rangle\;,\\
\label{boson}
&=&{\hbar n_s\over m}\;,
\end{eqnarray}
and $n_s$ the boson superfluid density. The well-known identification
of physical parameters under the boson--vortex mapping is given by
\begin{eqnarray}
\hbar&\leftrightarrow&k_BT\;,\\
m&\leftrightarrow&\tilde{\epsilon}_1\;,\\
\hbar\beta&\leftrightarrow&L\;,
\end{eqnarray}
with $m$ the boson mass, $\tilde{\epsilon}_1$ the core energy per unit
of length of a single vortex line, $\beta$ the inverse boson
temperature and $L$ the vortex sample thickness. Utilizing this
mapping, together with the approximate local expression for the tilt
modulus $c_{44}=n_0\tilde{\epsilon}_1$, we reassuringly find that the
defect-free flux-line lattice (in which vortex lines do not entangle
and the sample exhibits longitudinal superconductivity) corresponds to
the ``normal'' boson crystal with vanishing superfluid density
$n_s=0$. The entangled flux-line liquid, on the other hand,
corresponds to the superfluid phase of bosons with $n_s=n_0$, and does
not exhibit longitudinal superconductivity.  We stress, however, that
the mapping described above, as well as Eqs. \ref{winding} and
\ref{boson}, only apply to a model of the vortices that neglects the
nonlocality of the intervortex interaction along the field ($z$)
direction.\cite{pbmcm} A more general approach to the derivation of
vortex liquid hydrodynamics, directly from a ``kinetic theory'' of
interacting flux lines was developed in Ref.\onlinecite{RFhydro}.
\begin{figure}[bth]
{\centering
\setlength{\unitlength}{1mm}
\begin{picture}(150,85)(0,0)
\put(-27,-44){\begin{picture}(150,85)(0,0) 
\includegraphics{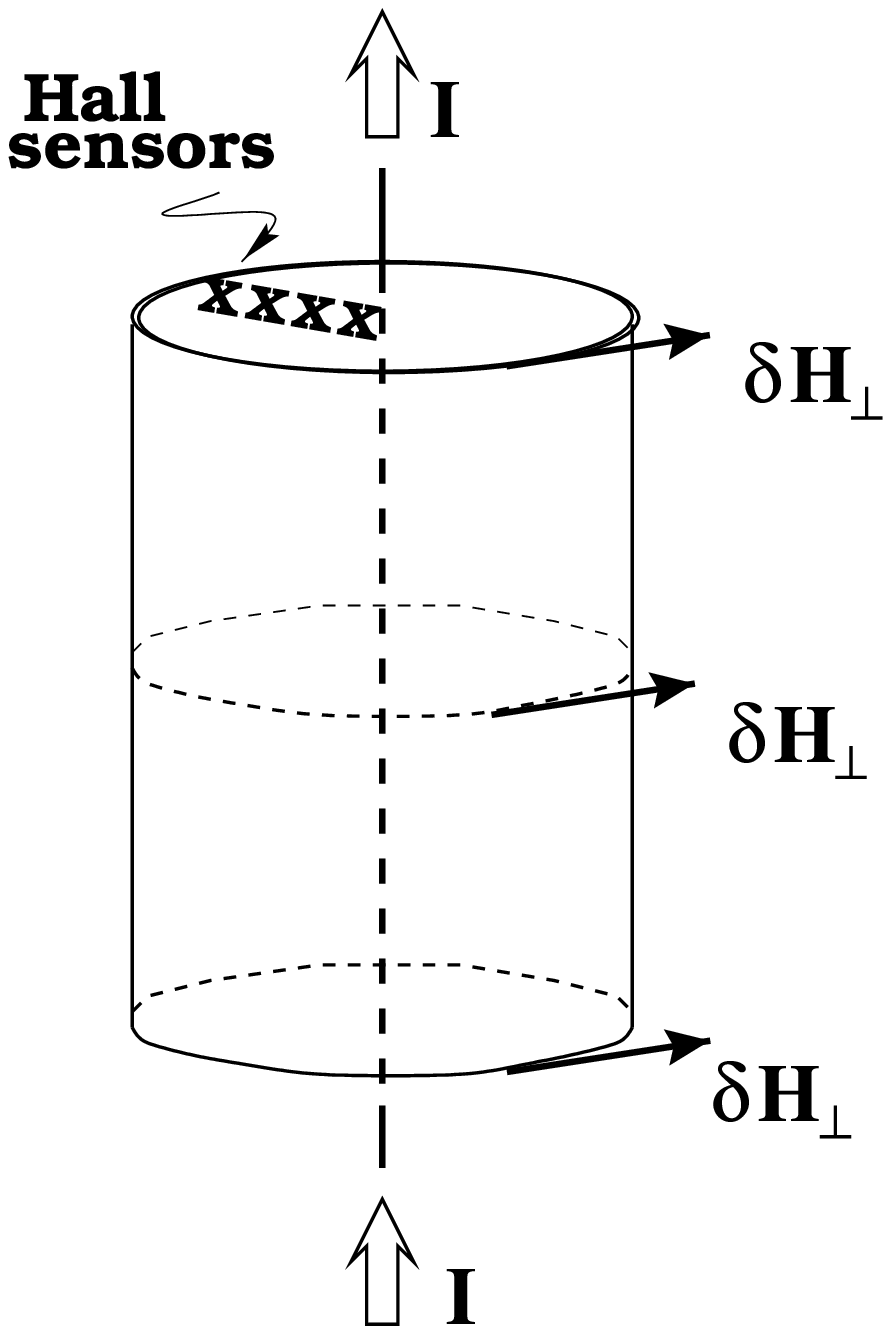}
\end{picture}}
\end{picture}}
Fig.2.{Sketch of an idealized experimental setup to
probe longitudinal superconductivity.}
\end{figure}

Figure 2 shows a sketch of an idealized experimental setup that could
be used to probe longitudinal superconductivity. An external field
${\bf H}_0$ is applied along the axis ($z$ in our coordinate system)
of a cylindrical sample and a uniform vortex state with flux lines
running along $z$ is set up. A current $I$ confined to a wire running
along the axis of the cylinder and producing an additional azimuthal
magnetic field $\delta{\bf H}_\perp({\bf r}_\perp)=2I\hat{\bf
z}\times{\bf r}_\perp/(r_\perp^2 c)$, which provides the shear
perturbation described above, can therefore probe the $q_z=0$ field
response of the vortex state.  The longitudinal superconductivity can
then be studied by measuring the induced azimuthal component
$B_\phi(r_\perp)$ of the local induction inside the sample.  This may
be possible by placing radially-directed row of Hall sensors at the
top of the sample.  In the defect-free crystal, which is a
longitudinal superconductor, we expect $B_\phi=0$ everywhere in the
bulk of the sample, deeper than the penetration length $\lambda$ from
the surface of the cylinder.  In contrast in the supersolid and
hexatic phases there will be a nonvanishing azimuthal response
$B_\phi(r_\perp)$ everywhere in the sample.

\section{Elastic properties of supersolids}
\label{supersolid}
\subsection{Model}
\label{model_supersolid}

As discussed in the Introduction, we expect that upon increasing the
temperature and the external magnetic field, the defect-free vortex
crystal will undergo a thermodynamically sharp transition into a
vortex supersolid, characterized by the coexistence of crystalline
order and a finite equilibrium density of vacancy and interstitial
defects.\cite{fnf} Our goal here is to develop a continuum description
of the long wavelength elastic properties of such a supersolid phase.

Once vacancy and interstitial line defects proliferate in the
supersolid, their positions and orientations represent new and
important low-energy degrees of freedom, independent of the lattice
displacements characterized by the field $\bf u({\bf r})$. At finite
temperature, these defects will lead to fluctuations in the local
induction, independent, but energetically coupled to, local elastic
strains.  To incorporate such defect fluctuations, we adapt the
hydrodynamic methods developed for vortex lines,\cite{MNhydro,RFhydro}
to the hydrodynamics of the vacancy and interstitial line-liquid.

At finite density, i.e., within the vortex supersolid phase, the
low-energy vacancy and interstitial configurations can be parameterized
by  $z$-{\em directed} conformations $({\bf r}_n^a(z),z)$, (with
$a=i,v$ denoting interstitial and vacancy, respectively) as they
traverse the sample along the direction of the applied field, in close
analogy to vortex lines themselves.  Long wavelength properties of a
gas of $N_v$ vacancies and $N_i$ interstitials can then be described
in terms of a net areal density of defects,
\begin{equation}
\label{nd}
n_d({\bf r}_\perp,z)=n_i({\bf r}_\perp,z)-n_v({\bf r}_\perp,z)\;,
\end{equation}
and a two-dimensional tilt vector field,
\begin{equation}
\label{td}
{\bf t}_d({\bf r}_\perp,z)={\bf t}_i({\bf r}_\perp,z)-
{\bf t}_v({\bf r}_\perp,z)\;,
\end{equation}
where 
\begin{equation}
\label{density}
n_a({\bf r}_\perp,z)=\sum_{n=1}^{N_a}\delta^{(2)}({\bf r}_\perp-{\bf
r}_n^a(z))\;,
\end{equation}
and
\begin{equation}
\label{tangent}
{\bf t}_a({\bf r}_\perp,z)=\sum_{n=1}^{N_a}
{\partial{\bf r}^a_n(z)\over\partial z}
\delta^{(2)}({\bf r}_\perp-r_n^a(z))\;.
\end{equation}
Here $n_d({\bf r}_\perp,z)$
represents the net number of defect lines crossing a unit area
perpendicular to the field direction, while $t_{di}({\bf r}_\perp,z)$
is the net number of defect lines crossing a unit area normal to the
$i$-th direction, with $i=x,y$.  Since the interaction among vacancy and
interstitial defects is short-ranged, we expect that a liquid of such
defects be characterized by a compressibility (inverse defect bulk
modulus), $\chi$, and a finite tilt modulus, $K$. The corresponding
long-wavelength free energy functional is therefore given by
\begin{equation}
\label{freed}
F_d={1\over 2 n_0^2}\int d{\bf r}\Big[\chi^{-1}(\delta n_d)^2 +K({\bf
t}_d)^2\Big]\;,
\end{equation}
with $\delta n_d=n_d-n_d^0$, and $n_d^0$ the mean net defect density
in equilibrium.

In the presence of these defects, fluctuations in the local magnetic
induction (or in the corresponding {\em flux}-line density $n({\bf
r})$ and {\em flux} tangent field ${\bf t}({\bf r})$) can be brought
about by changes in both the local lattice strains $w_{\alpha j}$ and
the defect densities $n_d$ and ${\bf t}_d$, as
\begin{equation}
\label{deltan}
\delta n = -n_0 w_{ii}+\delta n_d\;,
\end{equation}
and
\begin{equation}
\label{deltat}
t_i=n_0w_{zi}+t_{di}\;.
\end{equation}
In this case the continuity equation (\ref{continuity}), arising from
$\bbox{\nabla}\cdot{\bf B}=0$ yields the {\em nontrivial} constraint
\begin{equation}
\label{continuityd}
\partial_z\delta n_d+\bbox{\nabla}_\perp\cdot{\bf t}_d=0\;,
\end{equation}
that defect lines cannot start nor stop inside the sample.  The {\em
elastic} strain drops out from Eq.\ref{continuityd}, as it identically
satisfies the constraint due to the single-valuedness of the
displacement field ${\bf u}({\bf r})$.

Since motion of defects microscopically corresponds to hopping of
vortex lines, we expect an energetic coupling between fluctuations in
the density and orientation of defects and the elastic strain
field. The lowest order coupling allowed by symmetry corresponds to
the following interaction part of the free energy functional
\begin{equation}
\label{freeint}
F_{int}={1\over n_0}\int d{\bf r}
\Big[\gamma\delta n_d\bbox{\nabla}\cdot{\bf u}
+\gamma'{\bf t}_d\cdot\partial_z{\bf u}\Big]\;,
\end{equation}
where $\gamma$ and $\gamma'$ are positive phenomenological coupling
constants with dimensions of elastic moduli. Ignoring a weak coupling
to fluctuations in the local temperature, the free energy functionals
$F_{int}$ and $F_d$, together with the elastic part $F_{\rm latt}$,
Eq.\ref{freeel}, 
completely determine the long-scale elastic properties of the
vortex supersolid.

It is important to note that the parameters $c_{11}, c_{66}, c_{44},
\gamma, \gamma', \chi$, and $K$, appearing in our model, are 
functions of the {\em mean} net defect density $n_d^0$,\cite{ClareYu}
which, within the supersolid phase, can in principle be determined
through detailed microscopic calculations of the type presented in
Ref.\onlinecite{fnf}. At the vortex-crystal to supersolid transition,
we expect these parameters to display a nonanalytic behavior as a
function of the distance from the transition, $|T-T_{xss}|$ (where
$T_{xss}$ denotes the crystal-to-supersolid transition temperature),
of the form illustrated for $c_{66}(n_d^0(T))$ in Fig.3.
\begin{figure}[bth]
{\centering
\setlength{\unitlength}{1mm}
\begin{picture}(150,70)(0,0)
\put(-3,-25){\begin{picture}(150,70)(0,0) 
\includegraphics{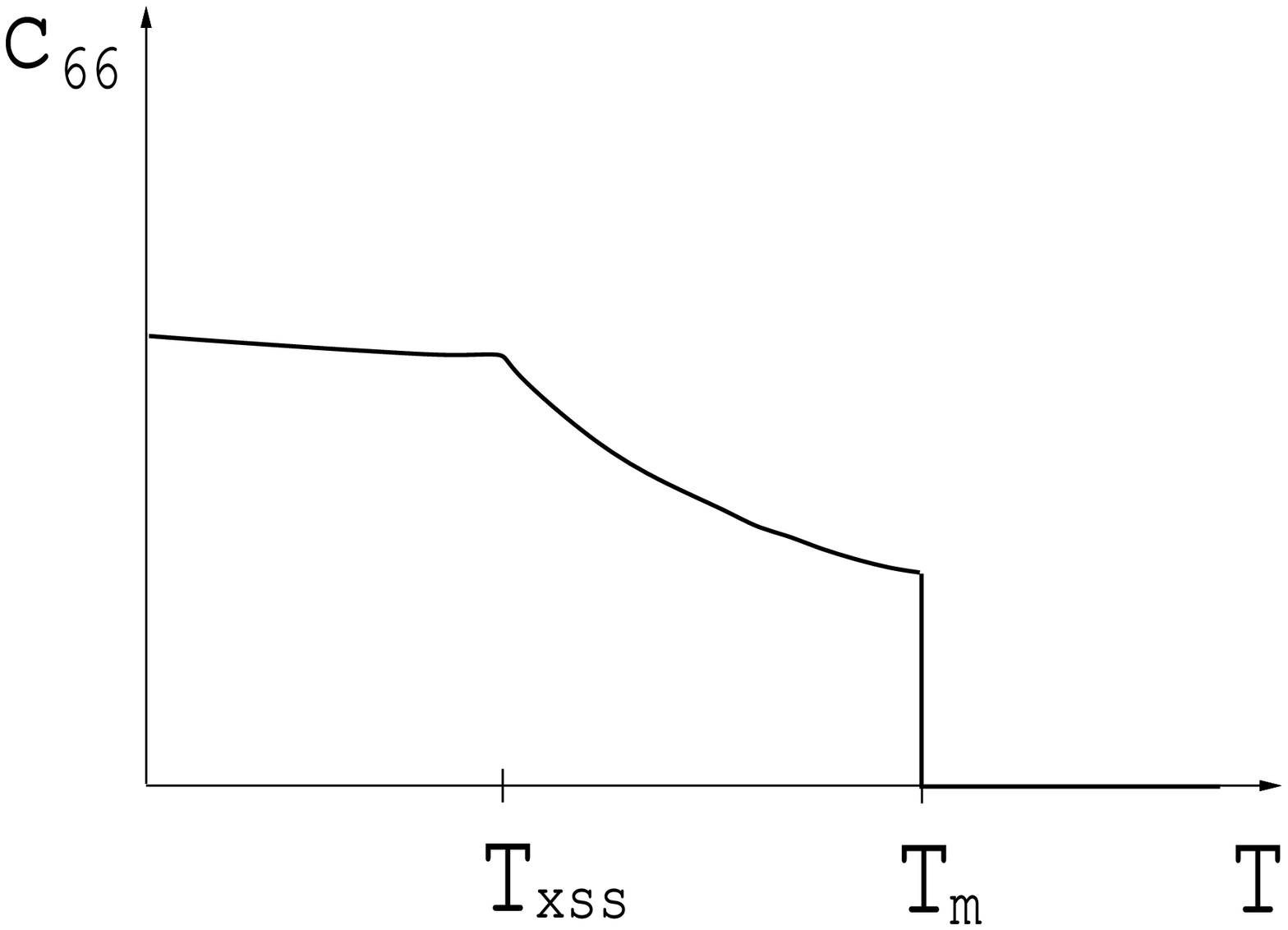}
\end{picture}}
\end{picture}}
Fig.3.{ A sketch illustrating the behavior of the elastic shear modulus
$c_{66}$ near the vortex crystal to vortex supersolid transition. The
proliferation of vacancies and interstitials is responsible for the
softening of the effective shear modulus inside the supersolid phase.}
\end{figure}

In addition to these mean-field effects, the coupling of the elastic
degrees of freedom to the {\em fluctuations} in defect density (around
the average $n_d^0$) and orientation field, Eq.\ref{freeint}, yields
further renormalization of the elastic constants, study of which is in
part the focus of our work here.

\subsection{Correlation and response functions}
\label{CRsupersolid_functions}

The elastic properties of supersolids can be characterized in a number of
distinct ways, reflecting a variety of experimental probes that couple
differently to the supersolid degrees of freedom. The simplest of these,
from the theoretical and experimental points of view, is the extension
of the equal-time equilibrium correlation functions $S({\bf q})$ and
$T_{ij}({\bf q})$ defined by Eqs.\ref{Sq} and \ref{Tq}, respectively.
They directly measure fluctuations in the local magnetic induction
${\bf B}({\bf r})$, related to the fluctuations in the {\em total}
vortex-line number and tilt densities, via relations (\ref{Bz}) and
(\ref{Bperp}). These latter quantities are determined by {\em both}
the local elastic strain and the defect density fields through
Eqs.\ref{deltan} and \ref{deltat}, and lead to
\begin{eqnarray}
S^{ss}({\bf q})&=&{1\over V}
\langle|n_0 w_{ii}({\bf q})-\delta n_d({\bf q})|^2\rangle
\;,\label{Sqss}\\
T_{ij}^{ss}({\bf q})&=&{1\over V}
\langle(n_0 w_{zi}({\bf q})+t_{di}({\bf q}))
(n_0 w_{zj}(-{\bf q})+t_{dj}(-{\bf q}))\rangle\;.\label{Tqss}\nonumber\\
\end{eqnarray}

In the above, the brackets denote a thermal average evaluated with the
total {\em supersolid} free energy functional which includes both the
elastic and the defects degrees of freedom, given by
\begin{equation}
\label{Fss}
F_{ss}=F_{\rm latt}+F_d+F_{int}\;.
\end{equation}
To ensure the condition of $\bbox{\nabla}\cdot{\bf B}=0$, these
averages must be carried out under the nontrivial constraint that
defect lines do not start nor stop inside the sample, given in
Eq.\ref{continuityd}.  

Utilizing this constraint to explicitly eliminate $t_d^L$ in favor of
$\delta n_d$ and re-expressing the strain tensor $w_{ij}$ in terms of
the longitudinal and transverse single-valued lattice displacements
\begin{eqnarray}
u_L({\bf q})=\hat{\bf q}_\perp\cdot{\bf u}({\bf q})\;,
\label{uL}\\
u_T({\bf q})=(\hat{\bf z}\times\hat{\bf q}_\perp)\cdot{\bf u}({\bf q})\;,
\label{uT}
\end{eqnarray}
which are the independent finite wavevector elastic degrees of freedom
in the bulk, we obtain the total free energy characterizing the
supersolid
\begin{eqnarray}
F_{ss}&=&\int {d^3{\bf q}\over(2\pi)^3}\Bigg\{
{1\over2}\Gamma_T({\bf q})|u_T({\bf q})|^2+
{1\over2}\Gamma_L({\bf q})|u_L({\bf q})|^2\nonumber\\
&+&{1\over 2n_0^2}\Big[
\left(\chi^{-1}+K {q_z^2\over q_\perp^2}\right)|\delta n_d({\bf q})|^2
+K |t_d^T({\bf q})|^2\Big]\nonumber\\
&+&{i\over n_0}\Big[
\left(\gamma q_\perp-
\gamma'{q_z^2\over q_\perp}\right)\delta n_d({\bf q})u_L(-{\bf q})\nonumber\\
&+&\gamma' q_z t_d^T({\bf q}) u_T(-{\bf q})\Big]\Bigg\}\;.\nonumber\\
\label{Fss_long}
\end{eqnarray}
In the above we have defined the transverse and longitudinal
wavevector-dependent stiffnesses
\begin{eqnarray}
\Gamma_T({\bf q})&=&c_{66}q_\perp^2+c_{44}q_z^2\;,\label{stiffnessesT}\\
\Gamma_L({\bf q})&=&c_{11}q_\perp^2+c_{44}q_z^2\;.\label{stiffnessesL}
\end{eqnarray}
and not surprisingly found that the transverse and the longitudinal
degrees of freedom decouple.  After re-expressing the $S({\bf q})$ and
$T_T({\bf q})$ in Eqs.\ref{Sqss} and \ref{Tqss} in terms of these same
independent degrees of freedom, these correlation can be easily
computed by inverting the corresponding $2\times2$ matrices that can
be read off from the expression for $F_{ss}$, Eq.\ref{Fss_long}.
For the structure function of the supersolid we thereby obtain
\end{multicols}
\begin{eqnarray}
\label{Sqssres}
S^{ss}(q_\perp,q_z)&=&{1\over V}
\langle|-i n_0 q_\perp u_L({\bf q})+\delta n_d({\bf q})|^2\rangle\;,
\nonumber\\
&=&{n_0^2k_BTq_\perp^2}{(c_{11}q_\perp^2+c_{44}q_z^2)
+(\chi^{-1}q_\perp^2+Kq_z^2) +2(\gamma q_\perp^2-\gamma'q_z^2)\over
(c_{11}q_\perp^2+c_{44}q_z^2)(\chi^{-1}q_\perp^2+Kq_z^2) -(\gamma
q_\perp^2-\gamma'q_z^2)^2}\;.
\end{eqnarray}
Similarly, the transverse part of the tilt correlation function is
given by
\begin{eqnarray}
\label{Ttqssres}
T_T^{ss}(q_\perp,q_z)&=&{1\over V}
\langle|i n_0 q_z u_T({\bf q})+t_{d}^T({\bf q})|^2\rangle\;,\nonumber\\
&=&{k_B T n_0^2\over K} + k_B T n_0^2
{q_z^2(1-\gamma'/K)^2\over q_\perp^2 c_{66}+q_z^2(c_{44}-\gamma'^2/K)}.
\end{eqnarray}
\begin{multicols}{2}

It is convenient to define defect-renormalized elastic constants that
characterize the effective elastic properties of supersolids. One
choice of such a definition, that naturally arises in experiments
which measure fluctuations in the {\em total} magnetic induction ${\bf
B}$ (e.g., neutron scattering experiments), is in terms of the static
correlation functions given in Eqs.\ref{Sqssres} and \ref{Ttqssres},
identified with their defect-free lattice counterparts, Eqs.\ref{Sq0}
and \ref{Tq0}
\begin{equation}
\label{c11def}
{n_0^2k_BT\over c_{11}^R(q_\perp)}\equiv S^{ss}(q_\perp, q_z=0)\;,
\end{equation}
and
\begin{equation}
\label{c44def}
{n_0^2k_BT\over c_{44}^R(q_z)}\equiv T_T^{ss}(q_\perp=0, q_z)\;.
\end{equation}
Similarly, the effective shear modulus $c_{66}^R$ of the supersolid
can be defined in terms of another equilibrium equal-time correlation
function
\begin{eqnarray}
\label{c66def}
{k_BT\over c_{66}^R(q_\perp)}&\equiv & 
V^{-1}\big[\langle w_{ij}({\bf q})w_{ij}(-{\bf q})\rangle
-\langle w_{ii}({\bf q})w_{jj}(-{\bf q})\rangle\big]\Big|_{q_z=0}\;,
\nonumber\\
&=&V^{-1}q_\perp^2P^T_{ij}({\bf\hat{q}_\perp})
\langle u_i({\bf q})u_j(-{\bf q})\rangle|_{q_z=0}\;,\nonumber\\
&=&V^{-1}q_\perp^2\langle|u_T({\bf q})|^2\rangle\Big|_{q_z=0}\;.
\end{eqnarray}
which can be easily evaluated by performing a Gaussian thermal average
with the free energy functional, Eq.\ref{Fss_long} in the Boltzmann weight.
Using Eqs.\ref{Sqssres} and \ref{Ttqssres} together with the definitions,
Eqs.\ref{c11def} and \ref{c44def}, we find
\begin{equation}
\label{c11ren}
c_{11}^R(q_\perp)={c_{11}\chi^{-1}-\gamma^2\over c_{11}+\chi^{-1}+2\gamma}\;,
\end{equation}
and
\begin{equation}
\label{c44ren}
c_{44}^R(q_z)={c_{44}K-\gamma'^2\over c_{44}+K-2\gamma'}\;.
\end{equation}
From these expressions we observe that for vanishing couplings
$\gamma=\gamma'=0$, the effective elastic moduli of the supersolid are
the corresponding moduli of the lattice and the liquid of defects,
added in ``parallel'' {---} a physically appealing result. For weak
coupling this implies that the effective bulk and tilt moduli of the
supersolid are always {\em smaller} than or equal to the minimum of
the corresponding moduli of the two subsystems, the lattice and the
defects. Hence we find that for $\gamma,\gamma'$ small compared to the
elastic moduli of the lattice and the defects, fluctuations of vacancy
and interstitial densities always {\em reduces} the effective
longitudinal and tilt moduli of the supersolid, relative to that of
the defect-free crystal.

The typical behavior of the supersolid bulk modulus $c_{11}^R$ for
$c_{11}\neq\chi^{-1}$ as a function of $\gamma$ is displayed in Fig.4
($c_{44}^R$ behaves similarly as function of $\gamma'$).  The figure
shows that the supersolid modulus grows (decreases) linearly with
positive (negative) coupling, at weak coupling. At an intermediate
{\em positive} value of $\gamma$, it reaches a maximum at the value of
the smallest of the moduli for the two subsystems.  In the strong
coupling regime (both positive and negative) the effective modulus
decreases, vanishes and even changes sign, indicating an instability
in the quadratic model of the supersolid.  In our model, this
instability is a signal of a (spurious) phase transition {\em within}
the supersolid phase, which in the case of the bulk modulus
$c_{11}^R$, corresponds to additional proliferation of defects and
change in the {\em lattice} density. However, as we show in Appendix
A, for a physical vortex supersolid, the elastic moduli
($c_{66},c_{11},c_{44},\chi,K,\gamma,\gamma'$) that appear in our
model are constrained to lie outside of this unstable range. In the
special degenerate case of $c_{11}=\chi^{-1}$, the effective bulk
modulus of the supersolid behaves as $c_{11}^R=(c_{11}+\gamma)/2$,
with an analogous result for the tilt modulus $c_{44}^R$.

The calculation of $c_{66}^R$, as defined in Eq.\ref{c66def}, shows
that, not surprisingly, vacancy and interstitial density {\em
fluctuations} do {\em not} renormalize the shear modulus,
and
\begin{equation}
c_{66}^R(q_\perp)=c_{66}\;. 
\end{equation}
In contrast to dislocations (to be considered in the next
section), vacancies and interstitials are unable to relieve a shear stress.
\begin{figure}[bth]
{\centering
\setlength{\unitlength}{1mm}
\begin{picture}(150,70)(0,0)
\put(-17,-42){\begin{picture}(150,70)(0,0) 
\includegraphics{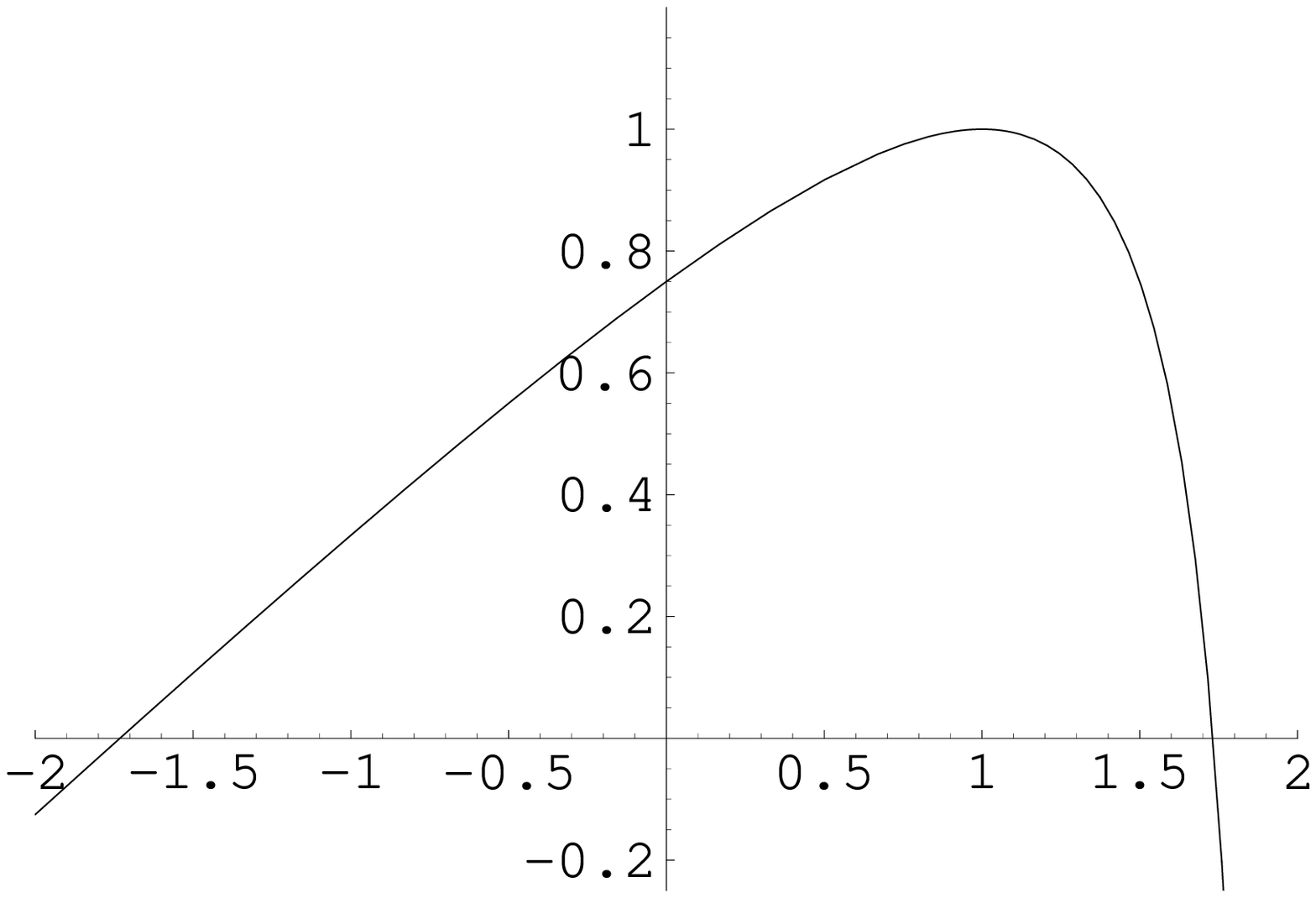}
\end{picture}}
\put(75,5){\LARGE{$\gamma$}}
\put(32,62){\LARGE{$c_{11}^R$}}
\end{picture}}
Fig.4.{The behavior of the supersolid bulk modulus as function of the
coupling $\gamma$ between the elastic and the defect degrees of freedom,
for $c_{11}=1<\chi^{-1}$.}
\end{figure}

The presence of vacancy and interstitial defects also alters the
response to a transverse magnetic field.  The long wavelength limit of
the transverse part of the tilt correlation function,
$T_T^{ss}(q_\perp,q_z)$, remains nonanalytic, as in a defect-free
crystal, and yields
\begin{eqnarray}
\label{chid}
& & \lim_{q_z\rightarrow 0}\chi_T^{ss}(q_\perp=0,q_z)
=-{1\over 4\pi}\Big[1-{B^2\over 4\pi c_{44}^R}\Big]\;,\\
\label{rhosd}
& & \lim_{q_\perp\rightarrow0}\chi_T^{ss}(q_\perp,q_z=0)=-{1\over 4\pi}
\Big[1-{B^2\over 4\pi K}\Big]\;.
\end{eqnarray}
Since, as discussed above, these defects soften the lattice by {\em
decreasing} both the longitudinal and tilt moduli the penetration of a
transverse field is {\em enhanced}, as seen from Eq.\ref{chid}.  A
more dramatic effect of the defects is the presence of the second term
on the right hand side of Eq.\ref{rhosd}. In the vortex supersolid
vacancy and interstitial defects allow flux-line wandering and
entanglement, and as a result there is no perfect screening of weak
transverse fields uniform along the $z$ axis, and consequently the
phase does not exhibit longitudinal superconductivity. The supersolid
is both crystalline ($c_{66}\neq0$) and entangled
($T_T^{ss}(q_\perp,q_z=0)\not=0$), as argued by Frey, et al.\cite{fnf}
and consistent with nonvanishing winding correlation function
\begin{eqnarray}
\label{bosonss}
\langle W^2\rangle &=&\lim_{q_\perp\rightarrow 0}T_T(q_\perp,q_z=0)\;,\\
&=&{k_B T n_0^2\over K}\;,
\end{eqnarray}

We stress that the definitions of the effective elastic moduli
characterizing a vortex supersolid is far from unique.  This is
related to the variety of experiments that probe {\em different}
physical properties of the vortex supersolid. Instead of the
correlation functions $S^{ss}({\bf q})$ and $T_{ij}^{ss}({\bf q})$
defined in Eqs.\ref{Sqss} and \ref{Tqss}, we could have instead
focused on
\begin{eqnarray}
\tilde{S}^{ss}({\bf q})&=&{n_0^2\over V}
\langle|w_{ii}({\bf q})|^2\rangle
\;,\label{tildeSqss}\\
\tilde{T}_{ij}^{ss}({\bf q})&=&{n_0^2\over V}
\langle  w_{zi}({\bf q})w_{zj}(-{\bf q})\rangle\;,
\label{tildeTqss}
\end{eqnarray}
which probe only fluctuations in the {\em lattice} positions
and {\em not} the {\em total vortex} density (related to magnetic
induction ${\bf B}$). These latter correlation functions are more
difficult to probe in a physical experiment, but can be
straightforwardly measured in a numerical simulation. Making the
identification between the effective elastic moduli of the supersolid and
the corresponding correlation functions, in analogy with
Eqs.\ref{c11def} and \ref{c44def}, we define
\begin{equation}
\label{tildec11def}
{n_0^2k_BT\over\tilde{c}_{11}^R(q_\perp)}\equiv\tilde{S}^{ss}(q_\perp,q_z=0)\;,
\end{equation}
and
\begin{equation}
\label{tildec44def}
{n_0^2k_BT\over\tilde{c}_{44}^R(q_z)}\equiv\tilde{T}_T^{ss}(q_\perp=0, q_z)\;.
\end{equation}
Simple computation of $\tilde{S}^{ss}({\bf q})$ and
$\tilde{T}_T^{ss}({\bf q})$, together with these definitions leads to 
\cite{notetilde}
\begin{eqnarray}
\tilde{c}_{11}^R&=&c_{11}-\gamma^2\chi\;,\label{tildec11ren}\\
\tilde{c}_{44}^R&=&c_{44}-{\gamma'^2\over K}\;,\label{tildec44ren}\\
\tilde{c}_{66}^R&=&c_{66}\;.\label{tildec66ren}
\end{eqnarray}

Another experimentally relevant way to probe elastic properties of
vortex supersolids is through the linear response to a constant stress
$\sigma_{\alpha j}$ applied at the boundaries of the system. For
simplicity, we confine our discussion here to a two-dimensional stress
$\sigma_{ij}$ applied to a boundary lying in the $xy$-plane.  To study
the response, we first need to decide to which physical quantity does
such stress couple. In a defect-free crystal, the answer is simple:
the stress $\sigma_{ij}$ couples to the lattice strain $u_{ij}$. In a
supersolid, there is, however, a number of possibilities, depending on
the nature of the experiment one seeks to describe (as was the case
with the correlation functions discussed above). In a real (as opposed
to a numerical) experiment the stress on the vortex lattice is
produced through an electromagnetic interaction and therefore couples
to the magnetic induction, which involves {\em both} the elastic
lattice strain tensor and the defect contribution. Arguments similar
to those found in Ref.\onlinecite{ZHN} indicate that the linear
response of the supersolid can be studied by adding to the free energy
$F_{ss}$ in Eq.\ref{Fss_long} a part due to the external stress,
\begin{equation}
F_\sigma=-\int d{\bf r}\,\sigma_{ij}(u_{ij}-
{1\over2n_0}\delta_{ij}\delta n_d)\;,
\label{Fsigma}
\end{equation}
and then by minimizing it with respect to the independent elastic strain
and defect degrees of freedom.

Before proceeding with the calculation, some remarks are in
order. Invoking the fluctuation-dissipation theorem, one might naively
conclude that such a static response function is identical to the
corresponding equal-time correlation function $S^{ss}({\bf q})$,
Eq.\ref{Sqss} that we studied above. However, this is in fact not the
case, in general, even for a defect-free crystal.  To understand this
difference, we consider for simplicity the case of a two-dimensional
crystal. Bulk correlation functions probe the fluctuations of the bulk
degrees of freedom at finite (albeit small) wavevector.  The {\em
elastic} degrees of freedom of the lattice or supersolid are the {\em
two} components of the lattice displacement (phonons), $u_x({\bf q})$
and $u_y({\bf q})$ (or equivalently $u_L({\bf q})$ and $u_T({\bf
q})$). Hence in the definition of $S^{ss}({\bf q})$, Eqs.\ref{Sqss}
and \ref{tildeSqss}, the average is over the {\em two} lattice
displacements (as well as over the defects), {\em not} the components
of the strain tensor, $w_{\alpha i}$, which can be written as
derivatives of the displacements and therefore are not independent
degrees of freedom.  In contrast, the response functions, that most
directly relate to experimental probes of elasticity, measure the
response to a (often {\em uniform}) stress applied at the {\em
boundary} of the solid.  The corresponding $q=0$, uniform deformations
are described by {\em three} zero-mode independent degrees of freedom,
corresponding to three macroscopic strains $u_{xx}$, $u_{yy}$ and
$u_{xy}$ that can be independently induced in a solid.  As a
consequence, for a supersolid (or a crystal in general) the equal-time
correlation functions differ from the corresponding static response
functions, as we now explicitly demonstrate.

Finally, we note that the situation is different in a liquid, where
dislocations have proliferated and act as additional degrees of
freedom. The proliferated dislocations lead to multi-valued lattice
displacements and thereby allow for a {\em transverse} part of the
strain $w_{ij}$, which (for the first index $i$), is forbidden in the
supersolid.  As a result, in a two-dimensional liquid, all {\em three}
components of the symmetric strain ($u_{xx}$, $u_{yy}$ and $u_{xy}$)
are independent degrees of freedom.

Returning to the derivation of the linear response to a perturbation described
by Eq. \ref{Fsigma}, we treat $u_{ij}$, $w_{zj}$ and the defect densities 
as independent degrees of freedom and minimize the total free energy,
$F_{tot}=F_{ss}+F_{\sigma}$, with respect to them, to obtain
\begin{eqnarray}
{\delta F_{tot}\over\delta u_{ij}}&=&2c_{66}u_{ij}+
\delta_{ij}(c_{11}-2c_{66})u_{kk}+
\delta_{ij}{\gamma\over n_0}\delta n_d-\sigma_{ij}\;,\nonumber\\
&=&0\;,\label{Fuij}\\
{\delta F_{tot}\over\delta w_{zj}}&=&c_{44}w_{zj}+
{\gamma'\over n_0}t_{dj}\;,\nonumber\\
&=&0\;,\label{Fwzj}\\
{\delta F_{tot}\over\delta n_d}&=&{\chi^{-1}\over n_0^2}\delta n_d +
{\gamma\over n_0}u_{kk}+{1\over 2n_0}\delta_{ij}\sigma_{ij}\;,\nonumber\\
&=&0\;,\label{Fnd}\\
{\delta F_{tot}\over\delta t_{di}}&=&K t_{di}+{\gamma'\over n_0}w_{zi}
\;,\nonumber\\
&=&0\;.\label{Ftdi}
\end{eqnarray}
Given that the applied stress $\sigma_{ij}$ is purely within the
$xy$-plane equations (\ref{Fwzj}) and (\ref{Ftdi})
give
\begin{eqnarray}
t_{di}&=&0\;,\\
w_{zi}&=&0\;,
\end{eqnarray}
and the response is $z$ independent. Solving the remaining equations
for the strain and the defect density, we find
\begin{eqnarray}
u_{kk}&=&{1+\gamma\chi\over2(c_{11}-c_{66}-\gamma^2\chi)}\,\sigma_{kk}\;,
\label{ukk_res}\\
{\delta n_{d}\over n_0}&=&-{\chi(c_{11}-c_{66}+\gamma)
\over2(c_{11}-c_{66}-\gamma^2\chi)}\,\sigma_{kk}\;,
\label{nd_res}
\end{eqnarray}
which, when used inside Eq.\ref{Fuij} give
\begin{equation}
u_{ij}=R_{ij,kl}^{ss}\sigma_{kl}\;,
\label{Rdefine}
\end{equation}
where $R_{ij,kl}^{ss}$ is the {\em uniform} static response function given
by
\begin{eqnarray}
R_{ij,kl}^{ss}&=&{(2c_{66}-c_{11})(1+\gamma\chi)+
2\gamma\chi(c_{11}-c_{66}+\gamma)\over
4c_{66}(c_{11}-c_{66}-\gamma^2\chi)}\,\delta_{ij}\delta_{kl}\nonumber\\
&+&{1\over2c_{66}}\delta_{ik}\delta_{jl}\;.
\label{Rssresult}
\end{eqnarray}

As a check we observe that for $\chi\rightarrow0$, which freezes out
the defects, $R_{ij,kl}^{ss}$ reduces to the well-known response
function for a defect-free crystal\cite{comment_stability}
\begin{eqnarray}
R_{ij,kl}^{crystal}&=&{(2c_{66}-c_{11})\over
4c_{66}(c_{11}-c_{66})}\,\delta_{ij}\delta_{kl}\nonumber\\
&+&{1\over2c_{66}}\delta_{ik}\delta_{jl}\;.
\label{Rcresult}
\end{eqnarray}
Consistent with our discussion above,
$R_{ij,kl}^{ss}$ differs from the corresponding $q_z=0$ equal-time
correlation function, computed with the lattice displacements as the
independent degrees of freedom. 

Using Eqs.\ref{Rdefine} and \ref{Rssresult} we can now compute the
response to any uniform stress, $\sigma_{ij}$. For example, a uniform
hydrostatic pressure $\delta p$ corresponds to
\begin{equation}
\sigma_{ij}=-\delta p \delta_{ij}\;,
\end{equation}
and we obtain the bulk modulus for the vortex supersolid as
\begin{eqnarray}
{1\over B_{ss}}&\equiv& -{1\over A}{\delta A\over\delta
p}\;,\nonumber\\
&=&{1\over n_0}{\delta n\over\delta p}\;,\nonumber\\
&=&-{u_{kk}\over\delta p}+{1\over n_0}{\delta n_d\over\delta p}\;,\nonumber\\
&=&{B+\chi^{-1}+2\gamma\over B\chi^{-1}-\gamma^2}\;,
\label{Bss}
\end{eqnarray}
where $B=c_{11}-c_{66}$ is the standard definition for a defect-free
crystal bulk modulus.  We note the similarity in form with the
correlation function $S^{ss}({\bf q}_\perp,q_z=0)$, that we used to
define $c_{11}^R$, Eq.\ref{c11def}. As discussed above, the
computation of the uniform static {\em response function} gives the
{\em bulk modulus}, $c_{11}-c_{66}$, while the equal-time correlation
function (at $q_z=0$) defines the longitudinal phonon modulus,
$c_{11}$.

As we observed above for the effective $c_{11}^R$ in Eq.\ref{c44ren},
here too in the absence of interaction between the defects and the
lattice ($\gamma=0$), the supersolid bulk modulus is simply determined
by the bulk moduli of the two systems, added in parallel,
$B_{ss}^{-1}=B^{-1}+\chi$.  

Significant caution must be applied in comparing expression
(\ref{Bss}) with experiments. If, for example, the compression is
performed on a time scale that is slow compared to the relaxation time
of the elastic degrees of freedom, but fast compared to that of the
defects (which, being a conserved ``charge'' density, relaxes only
diffusively), then the defects are effectively frozen on the time
scale of the experiment. In this case, corresponding to
$\chi=0$, only elastic degrees of freedom can respond
and the result for $B_{ss}$ is simply $B$ of the defect-free crystal.

The vortex supersolid shear modulus $c_{66}^{ss}$ can be probed by
applying a constant force normal to two opposite lateral boundaries of
a rectangular $xy$-crossection sample. The corresponding stress is given
by
\begin{equation}
\sigma_{ij}= \sigma\delta_{ix}\delta_{jx}\;.
\end{equation}
Such stress induces both the longitudinal ($u_L$) and the transverse
($u_T$) deformations of the lattice.  The shear modulus is defined in
terms of the normal strain, $u_{xx}-u_{yy}$, as
\begin{equation}
{1\over c_{66}^{ss}}=2{u_{yy}-u_{xx}\over \sigma}\;.
\end{equation}
Consistent with our earlier calculation that used equal-time
correlation function, 
we find here that
\begin{equation}
c_{66}^{ss}=c_{66}\,
\end{equation}
i.e., the shear modulus remains unrenormalized by fluctuations in
density of vacancies and interstitials.

\section{Dislocation Loops in a Vortex Supersolid: a Model of Vortex
Liquid}
\label{liquid}

We now turn to the description of a vortex liquid. This is a
disordered, dissipative state of the flux-line array, that results
from either a direct melting of a defect-free vortex crystal or
(possibly) a continuous melting\cite{continuous_melting} of a vortex
supersolid, discussed in previous sections.

Following ideas of Kosterlitz and Thouless,\cite{KT} extended in
Ref.\onlinecite{mcmdrn} to three-dimensional vortex systems, we
describe a flux-line liquid as a supersolid with a finite equilibrium
concentration of unbound dislocation loops and vacancy/interstitial
defect lines. Such an approach nicely complements the more
conventional hydrodynamic description of the vortex liquid studied in
Refs.\onlinecite{MNhydro,RFhydro}. While somewhat more involved, the
advantage of the approach taken here is that it provides a more direct
connection between the vortex liquid and vortex {\em ordered} phases
(crystals), in which the defects are bound, thereby presenting a
unified description. It also provides a valuable detailed {\em
``microscopic''} characterization of the distinction between vortex
liquids and solids.

The properties of a flux-line lattice in the presence of an unbound gas
of dislocation loops, but no vacancies nor interstitials, were studied
in Ref. \onlinecite{mcmdrn}. In the absence of vacancy and
interstitial defects, dislocation loops are constrained to lie in a
plane spanned by their Burger's vector and the $z$ axis.  These planar
dislocation loops can only relax applied stresses by ``gliding'' along
the $z$ axis. Clearly, once dislocation loops proliferate on all
scales, vacancies and interstitials will also unbind and {\em both} of
these defects will exist in the resulting bond-orientationally ordered
liquid.  The goal of this section is to incorporate
vacancies and interstitials into a complete description of a vortex
liquid. We will explicitly
demonstrate that a finite concentration of vacancy and interstitial
defects allows for ``climb-like'' distortions of dislocation loops,
which can 
move out of the $\hat{\bf z}$-$\bf b$ plane by
absorbing and emitting vacancies and interstitials.

\subsection{Model}
\label{model_liquid}

To construct a complete model of a vortex liquid we now proceed to
incorporate dislocation loops into the model of a supersolid,
presented in Sec.\ref{supersolid}. We do this by allowing multi-valued
lattice displacement fields, $\bf u$. The loop integral of $\bf u$,
\begin{equation}
\label{burger}
\oint du_i({\bf r}) =-b_i({\bf r})\;,
\end{equation}
enclosing a dislocation line, fails to close by a Burger's lattice
vector, $\bf b$.  The direction of integration around the contour is
that of a right-handed screw advancing parallel to a unit tangent
vector $\bbox \tau$ to the dislocation line.  The peculiarity of
dislocation in a lattice of $z$-directed lines is that while the
Burger's vector is two-dimensional and by definition lies in the $xy$
plane, the tangent $\bbox \tau$ to the defect line is a
three-dimensional vector.\cite{mcmdrn,nabarro} 

To study properties of the system on scales that are long compared to
the spacing between dislocation lines, we use a continuum
description. We consider a small hydrodynamic volume and introduce the
Burger's ``charge'' density {\em tensor} $\alpha_{\beta j}({\bf r})$,
whose integral over an open surface $S$, gives the total Burger's
vector of dislocation lines directed along the surface normal
$n_\beta$ and enclosed by a contour $C$ bounding the surface,
\begin{equation}
\label{distensor}
\int_S\alpha_{\beta j} n_\beta dA=\sum_nb_j^{(n)}\;.
\end{equation}
For a single dislocation line, directed along the tangent
$\tau_\beta$, with Burger's vector $b_i$, the defect density tensor is
given by $\alpha_{\beta i}=\tau_\beta b_i\delta^{(2)}(r_\perp)$. The
rectangular ($3\times2$) density tensor $\alpha_{\beta j}({\bf r})$ is
therefore a measure of the number of dislocation lines with Burger's
vector $b_j$ crossing a unit area normal to the dislocation tangent
$\tau_\beta$. We remind the reader that Roman letters $i,j,k,...$ are
used to denote indices that run only over the values $x$ and $y$, and
Greek letters $\alpha,\beta,\gamma,...$ are reserved for indices that
run over the three-dimensional set $x,y,z$.

By definition, dislocations in the $z$-directed line crystal must have
their Burger's vectors lie in the $xy$ plane. Consequently the
three-dimensional vector $\alpha_{zj}({\bf r})$, ($j=x,y$) describes
$z$-directed edge dislocations. Edge dislocations lying in the $xy$
plane are described by the antisymmetric part of the two-dimensional
tensor $\alpha_{ij}^\perp({\bf r})$, ($i=x,y$ and $j=x,y$, with
$\alpha_{\beta j}=(\alpha_{ij}^\perp,\alpha_{zj})$). In the {\em
absence} of vacancies and interstitials, however, these type of
$xy$-plane-directed edge dislocations correspond to a branching or
merging of flux lines, which necessarily involve fractional or double
flux-``charged'' vortex lines, both energetically forbidden. In
contrast, vacancies and interstitials allow for this type of edge
dislocations, as we will demonstrate below. Screw dislocations, on the
other hand always run normal to flux lines, i.e., they lie in the $xy$
plane, and lead to entanglement of the vortex lines.\cite{brandtent}
They are described by the {\em symmetric} part of
$\alpha_{ij}^\perp({\bf r})$.

By rewriting Eq.\ref{burger} in differential form and then averaging
the resulting equation over a hydrodynamic volume, containing many
dislocation lines, we relate the dislocation density tensor
$\alpha_{\beta j}({\bf r})$ to the local lattice strain $w_{\beta
j}({\bf r})$,
\begin{equation}
\label{diff}
\epsilon_{\alpha\beta\gamma}\partial_\beta w_{\gamma k}({\bf r})
=-\alpha_{\alpha k}({\bf r})\;.
\end{equation}
Hence in the presence of dislocations ($\alpha_{\beta j}\neq0$), the
strain tensor $w_{\beta j}$ contains a singular part that cannot be
written as a gradient ($\partial_\beta$) of a single-valued
displacement field $u_j$.  Finally, dislocation loops must either
close or terminate at sample boundaries.  This amounts to the
condition
\begin{equation}
\label{contdis}
\partial_\beta \alpha_{\beta j}({\bf r})=0\;.
\end{equation}

We now proceed to incorporate the dislocation degrees of
freedom into the model of the supersolid studied in
Sec.\ref{supersolid}. We first recall that the fluctuations of the
local magnetic induction $\delta{\bf B}=({\bf B}_\perp,\delta B_z)$
are related to the changes in density and orientation of flux lines,
with contributions from both local strains and vacancy and
interstitial defects. In the long wavelength limit these relations are
obtained by inserting $\delta n$ and ${\bf
t}$ given by Eqs.\ref{deltan} and \ref{deltat} in Eqs.\ref{Bz} and \ref{Bperp},
\begin{eqnarray}
\label{Bzd}
\delta B_z&=&\phi_0[-n_0w_{ii}({\bf r})+\delta n_d({\bf r})]\;,\\
\label{Bperpd}
\delta B_{\perp i}&=&\phi_0[n_0w_{zi}({\bf r})+t_{d i}({\bf r})]\;.
\end{eqnarray}
In the presence of dislocations,
the strain tensor $w_{\alpha j}$ is given by the sum of a regular,
longitudinal (on first, Greek index) part, defined in Eq.\ref{strain}
in terms of the derivatives of a single-valued displacement field
$u_j$, and a singular, transverse (on the first, Greek index) part,
$w_{\alpha i}^{\rm s}$, due to dislocations
\begin{equation}
\label{straintot}
w_{\alpha j}=\partial_\alpha u_j+w_{\alpha j}^{\rm s}\;.
\end{equation}

By imposing the $\bbox\nabla\cdot{\bf B}=0$ condition 
and using Eqs.\ref{Bzd} and \ref{Bperpd}, we obtain
\begin{equation}
\label{divB}
\partial_z\delta n_d+\bbox{\nabla}_\perp\cdot{\bf t}_d=
-n_0(-\partial_z w_{ii}^{\rm sing}+\partial_i w_{zi}^{\rm sing})\;.
\end{equation}
It is important to note that, in contrast to the defect-free crystal
and the supersolid, where the {\em elastic} part $w_{\beta
j}$ ($\partial_\beta u_j$) identically satisfies $\bbox\nabla\cdot{\bf
B}=0$, here this condition imposes a nontrivial constraint 
that couples dislocations and vacancy-interstitial
defects. This becomes apparent by using
Eq.\ref{diff} to eliminate $w_{\beta j}$ from the constraint in
favor of the dislocation density tensor, with the result
\begin{equation}
\label{contdv}
\partial_z\delta n_d+\bbox{\nabla}_\perp\cdot{\bf
t}_d=n_0\epsilon_{ij}\alpha_{ij}^\perp({\bf r})\;.
\end{equation}

This important condition is one of the main results of our work and is
an essential ingredient in the complete ``elastic'' description of the
vortex liquid state. We first note that in the supersolid phase, where
dislocation loops are bound and lattice displacements are
singe-valued, $\alpha_{\beta j}=0$, and the constraint reduces to the
vacancy-interstitial line continuity condition,
Eq.\ref{continuityd}. On the other hand, in a description of a vortex
liquid which ignores vacancies and interstitials it reduces to the
$\epsilon_{ij}\alpha_{ij}^\perp=0$ condition of
Ref.\onlinecite{mcmdrn}, enforcing the constraint, discussed above,
that in the absence of vacancies and interstitials, dislocation loops
are confined to lie in the plane defined by their Burger's vector and
the average external magnetic field.  In the presence of vacancies and
interstitials Eq.\ref{contdv} explicitly demonstrates that these
defects provide a mechanism by which the dislocation loop can
effectively climb out of this plane. They do this by emitting or
absorbing vacancy and interstitial line defects.  The allowance of
{\em nonplanar} dislocation loops in vortex supersolids and liquids
then removes the artificial condition of a vanishing antisymmetric
part of the dislocation density tensor and in general
$\epsilon_{ij}\alpha_{ij}^\perp\neq0$ in these phases. 
As anticipated above, this allows $xy$-plane-directed
{\em edge} dislocations to proliferate upon melting.

The above constraint, Eq.\ref{contdv}, is one way that the additional
degrees of freedom associated with proliferated dislocations enter the
description of the vortex liquid. Dislocations of course also
contribute directly through the ``elastic'' free energy in a way that
we now derive.

The free energy $F$ of a lattice with dislocations and
vacancy/interstitial defects is given by $F_{ss}=F_{\rm
latt}+F_d+F_{int}$, but with derivatives of the phonon field replaced
by the total strain tensor given in Eq. \ref{straintot}, i.e.,
\begin{eqnarray}
\label{freetot}
F=& &{1\over 2}\int d{\bf r}w_{\alpha i}C_{\alpha i\beta j}w_{\beta j}\\\nonumber
   & & +{1\over 2 n_0^2}\int d{\bf r}\big[\chi^{-1}(\delta n_d)^2
          +K({\bf t}_d)^2\big]\\\nonumber
   & & +{1\over n_0}\int d{\bf r}\big[\gamma w_{ii}\delta n_d
          +\gamma'w_{zi}t_{di}\big]\\\nonumber
   & & +F_{\rm core},
\end{eqnarray}
To make the notation more compact, the purely elastic part of the free
energy, given by the first term on the right hand side of
Eq. \ref{freetot}, has been written in terms of the elastic tensor,
\begin{eqnarray}
C_{\alpha i\beta j}=& &c_{66}(\delta_{\alpha\beta}-\delta_{\alpha z}\delta_{\beta z})\delta_{ij}
  +c_{66}\delta_{\alpha j}\delta_{\beta i}\nonumber\\
& &  +(c_{11}-2c_{66})\delta_{\alpha i}\delta_{\beta j}
  +c_{44}\delta_{\alpha z}\delta_{\beta z}\delta_{ij}\;.
\label{Caibj}
\end{eqnarray}
We have also added to the free energy of the defective lattice a term
$F_{\rm core}$ representing the core energy of the dislocations, given
by
\begin{equation}
\label{Fcore}
F_{\rm core}=\int d{\bf r} [E_e\alpha_{zi}^2+E_s\alpha_{ii}^2+E'_s\alpha_{ij}\alpha_{ij}
   +E'_e(\epsilon_{ij}\alpha_{ij})^2].
\end{equation}
The core energy $F_{\rm core}$ has been written on the basis of
symmetry considerations. It incorporates terms accounting for the edge
and screw dislocation core energies per unit length, $E_e$, $E'_e$,
$E_s$ and $E'_s$.  As discussed in Ref.\onlinecite{mcmdrn}, although
$E'_e=E'_s=0$ for a single dislocation line, nonzero values of $E'_e$
and $E'_s$ are required to describe short range interactions in the
hydrodynamic limit. The values for the core energies are estimated to
be $E_e\simeq c_{66}b^2$, $E_s\simeq
E'_s\simeq\sqrt{c_{66}c_{44}}\,b^2$ and
$E'_e\simeq\sqrt{c_{11}c_{44}}\,b^2$.\cite{mcmdrn}

While the elastic part ($\partial_\alpha u_i$) of the strain tensor
identically satisfies Eq.\ref{diff}, the singular part is found by
``inverting'' this equation in Fourier space. This gives
\begin{equation}
\label{wsolution}
w_{\alpha j}^{\rm s}({\bf q})=-{i\over
q^2}\epsilon_{\alpha\beta\gamma}q_\beta\alpha_{\gamma j}({\bf q})
+iq_\alpha\psi_j({\bf q})\;,
\end{equation}
where $\psi_j$ is an arbitrary function, reflecting the fact that the
solution to Eq.\ref{diff} is only determined up to an arbitrary
longitudinal part. This is analogous to the gauge freedom that appears in
electromagnetism (or other gauge theories) when, for example, Maxwell
equation ${\bbox\nabla}\cdot{\bf E}=\rho$ is solved for the electric
field $\bf E$ in terms of the charge density $\rho$. 

One convenient and natural choice of $\psi_j$ is obtained by requiring
that the corresponding $w_{\alpha j}^{\rm s}$ 
minimizes the total free energy $F$, Eq.\ref{freetot}, with respect to
lattice displacements. Such a choice is mathematically convenient for
computing thermodynamic averages, because it removes all couplings between
the phonons $u_j$ and the defect degrees of freedom.
Physically, this choice of $w_{\alpha j}^{\rm s}$
corresponds to elastically equilibrated defects.

To compute $\psi_i$ corresponding to this convenient gauge choice, we
insert $w_{\alpha i}$, Eq.\ref{straintot} into $F$, Eq.\ref{freetot},
and require that the {\em linear} terms in $\bf u$ vanish, which of
course is equivalent to the requirement that $w_{\alpha i}^{\rm s}$
minimizes $F$. The resulting Euler-Lagrange equation is given by
\begin{equation}
{\delta F\over \delta u_i}=-\partial_\alpha\sigma_{\alpha i}
-{\gamma\over n_0}\partial_i\delta n_d
-{\gamma'\over n_0}\partial_z t_{di}=0,
\end{equation}
where 
\begin{equation}
\label{stress}
\sigma_{\alpha i}=C_{\alpha i\beta j}w_{\beta j}
\end{equation}
is the corresponding stress tensor.
The solution is more conveniently written in Fourier space, where it
is given by
\begin{eqnarray}
\label{psisolution}
\psi_i({\bf q})=& &{1\over q^2}\big(A^{-1}\big)_{ij}q_\alpha C_{\alpha j\beta k}\epsilon_{\beta\gamma\eta}
   q_\gamma \alpha_{\eta k}({\bf q})\\\nonumber
& & +{i\over n_0}\big(A^{-1}\big)_{ij}\big[\gamma q_j\delta n_d({\bf q})
+\gamma' q_z t_{dj}({\bf q})\big].
\end{eqnarray}
Here $A^{-1}$ is the inverse of a $2\times 2$ matrix $A$, with
$A_{ij}=q_\alpha C_{\alpha i\beta j}q_\beta$. Its elements are given by,
\begin{equation}
\big(A^{-1}\big)_{ij}={1\over\Gamma_T({\bf q})}\bigg[\delta_{ij}-
 {(c_{11}-c_{66})\over\Gamma_L({\bf q})}q_{\perp i} q_{\perp j}\bigg],
\end{equation}
where $\Gamma_T({\bf q})$ and $\Gamma_L({\bf q})$ are the transverse and
longitudinal elastic elastic stiffnesses, defined in
Eqs. \ref{stiffnessesT},\ \ref{stiffnessesL}.

With this choice of $\psi_i$, by construction, the total free energy
breaks up into two parts
\begin{equation}
\label{FelFdef}
F=F_{\rm latt}+F_{\rm def},
\end{equation}
where $F_{\rm latt}$ is the defect-free elastic part of the free
energy given in Eq. \ref{freeel} and rewritten here for convenience in Fourier space,
\begin{equation}
F_{\rm latt}={1\over2}\int {d^3{\bf q}\over(2\pi)^3}\Big[
\Gamma_T({\bf q})|u_T({\bf q})|^2+
\Gamma_L({\bf q})|u_L({\bf q})|^2\Big]\;.
\label{Felphonon}
\end{equation}
This involves only phonons degrees of freedom, as for a defect-free crystal.  The
second, defect part, $F_{\rm def}$, is the free energy of an interacting gas
of dislocation loops and vacancy-interstitial line-defects.  It is
given by
\begin{eqnarray}
\label{freedef}
F_{\rm def}=& &{1\over 2}\int d{\bf r}w^s_{\alpha i}C_{\alpha i\beta j}w^s_{\beta j}\\\nonumber
   & & +{1\over 2 n_0^2}\int d{\bf r}\big[\chi^{-1}(\delta n_d)^2
          +K({\bf t}_d)^2\big]\\\nonumber
   & & +{1\over n_0}\int d{\bf r}\big[\gamma w^s_{ii}\delta n_d+\gamma'w^s_{zi}t_{di}\big],
\end{eqnarray}
where $w_{\alpha i}^s$ is the singular part of the strain tensor, given by
Eqs. \ref{wsolution} and \ref{psisolution}.
This defect free energy must be supplemented with the
continuity conditions
Eq.\ref{contdv} and Eq.\ref{contdis} for the vacancy/interstitial defect lines
and the dislocation lines, respectively.  The 
constraint (\ref{contdv})  can be directly incorporated into the free
energy by using it to explicitly eliminate the {\em longitudinal} part
of the vacancy-interstitial defects tangent vector,
\begin{eqnarray}
t_{d}^L({\bf q})&\equiv&{\bf \hat{q}}_\perp\cdot{\bf t}_{d}({\bf
q})\;,\label{tdL}\\
&=&-{q_z\over q_\perp}\delta n_d({\bf q})-{i\over q_\perp}
n_0\epsilon_{ij}\alpha_{ij}^\perp({\bf q})\;,
\end{eqnarray}
in favor of the defect density $\delta n_d({\bf q})$ and the
antisymmetric part of the dislocation density tensor
$\alpha_{ij}^\perp({\bf q})$. 
Finally, using Eqs. \ref{wsolution} and \ref{psisolution} to eliminate the singular strain
field in terms of the defect degrees of freedom in Eq. \ref{freedef},
we obtain
\end{multicols}
\begin{eqnarray}
\label{freealpha}
F_{\rm def}&=&
  {1\over 2}\int{d{\bf q}\over (2\pi)^3}\bigg\{
     \tilde{R}_{\mu i\nu j}({\bf q})
     \alpha_{\mu i}({\bf q})\alpha_{\nu j}(-{\bf q}) 
  + A({\bf q})\delta n_d({\bf q})  \delta n_d(-{\bf q}) 
  + C({\bf q})t_{d}^T({\bf q})t_{d}^T(-{\bf q})\nonumber\\ 
  &+&iD_{\mu i}({\bf q})\Big[\alpha_{\mu i}(-{\bf q})\delta n_d({\bf q})
          -\alpha_{\mu i}({\bf q})\delta n_d(-{\bf q})\Big]
  +iG_{\mu i}({\bf q})\Big[\alpha_{\mu i}(-{\bf q})t_{d}^T({\bf q})
          -\alpha_{\mu i}({\bf q})t_{d}^T(-{\bf q})\Big]
       \bigg\}\;,
\end{eqnarray}
where ${\bf t}_d^T({\bf q})={\bf t}_d({\bf q})-{\bf \hat{q}}_\perp
t^L_{d}({\bf q})$ is the transverse part of the defect tangent vector.
\begin{multicols}{2}
The kernels $\tilde{R}_{\mu i,\nu j}({\bf q})$, $A({\bf q})$, $C({\bf q})$,
$D_{\mu i}({\bf q})$ and $G_{\mu i}({\bf q})$ depend in a complicated
way on the elastic moduli ($c_{66},c_{11},c_{44}$) of the lattice, and
on wavevectors $q_\perp$ and $q_z$, and are given explicitly in
Appendix B.

\subsection{Correlations and response functions}
\label{CRliquid_functions}

In this section we evaluate the renormalization of the elastic
constants of the flux-line lattice due to dislocations and
vacancy/interstitial defects.  The renormalized elastic constants are
defined by Eqs. (\ref{c11def}-\ref{c66def}), but with the
understanding that the correlation functions are now those of a
lattice with an equilibrated concentration of defects and
dislocations. This means that the structure function and the tilt
correlation function are formally given in terms of the strain tensor
and the defect fields by the same expressions \ref{Sqss} and
\ref{Tqss} used for the supersolid, but the strain tensor $w_{\alpha
i}$ is now the total strain given in Eq. \ref{straintot}, including
the singular part.  The same holds for the correlation function that
determined the shear modulus defined on the first line of
Eq. \ref{c66def}.  The brackets in these correlation functions now
denote a thermal average with the free energy $F_{\rm latt}+F_{\rm
def}$, with $F_{\rm def}$ given by Eq. \ref{freealpha}.  The average
is carried out by integrating over all configurations of dislocations,
described by the components of the dislocation density tensor,
$\alpha_{\beta i}$, and vacancy/interstitial lines, described by the
defects density, $\delta n_d$, and tilt field, $t_{d}^T$.  The
integration must be done subject to the constraint that dislocation
lines are closed, given by Eq. \ref{contdis}. The ``continuity''
constraint for defect lines, expressed by Eq. \ref{contdv}, has
already been incorporated into the free energy $F_{\rm def}$.  The
computation of correlation functions is conceptually simple but
technically quite involved and was carried out using Mathematica
symbolic manipulator program. Only the results will be given here.

The full expression for the density, tilt field, and other correlation
functions are too ``horrifying'' to be shown here, and we therefore only
display their long wavelength limits, which determine the renormalized
elastic constants, according to Eqs. (\ref{c11def}-\ref{c66def}).

The structure factor vanishes as $q_\perp\rightarrow 0$, as required
by the density sum rule. The renormalized longitudinal modulus as
defined by Eq. (\ref{c11def}) is given by,
\end{multicols}
\begin{equation}
\label{c11result}
{1\over c_{11}^R(q_\perp)}={1\over c_{11}}\bigg\{1
   +{c_{11}[2c_{66}(c_{11}-c_{66})+c_{11}E_eq_\perp^2]+2c_{66}^2\chi^{-1}
   +\gamma(\gamma+2c_{11})(2c_{66}+E_eq_\perp^2)
   \over
   \chi^{-1}[2c_{66}(c_{11}-c_{66})+c_{11}E_eq_\perp^2]
   -\gamma^2(2c_{66}+E_eq_\perp^2)}\bigg\}.
\end{equation}
\begin{multicols}{2}
The renormalized tilt modulus defined by Eq. (\ref{c44def}) is 
\begin{equation}
\label{c44result}
{1\over c_{44}^R(q_z)}=\lim_{q_z\rightarrow 0}T_T(q_z,q_\perp=0)=
{c_{44}+K-2\gamma'\over c_{44}K-\gamma '^2}\;.
\end{equation}
We find that in the long wavelength limit both $c_{11}^R$ and
$c_{44}^R$ are identical to the elastic constants of a lattice with
only vacancy and interstitial defects.  In other words, somewhat
surprisingly, the coupling of dislocations to vacancy/interstitial
defect lines does {\em not} yield any {\em additional} renormalization
of the tilt and compressional moduli, even if it does make it possible
for dislocations to relax stresses by climbing out of the $({\bf
b},\hat{\bf z})$ plane and allows for (otherwise forbiddingly costly)
$xy$ directed edge dislocations.  Finally, the renormalized shear
modulus is
\end{multicols}
\begin{equation}
\label{c66result}
{1\over c_{66}^R(q_\perp)}={1\over c_{66}}+{1\over E_eq_\perp^2}
 +{\chi^{-1}c_{11}(c_{11}-2c_{66})-\gamma^2(c_{11}-4c_{66}) \over
   c_{11}\Big[\chi^{-1}\big(2c_{66}(c_{11}-c_{66})+c_{11}E_eq_\perp^2\big)
   -\gamma^2(2c_{66}+E_eq_\perp^2)\Big]}.
\end{equation}
\begin{multicols}{2}
\noindent Dislocations renormalize the long wavelength
($q\rightarrow0$) shear modulus to zero, yielding liquid-like response
to shear stresses.

We now discuss various limiting cases for our results.  In the absence
of coupling between the gas of vacancy/interstitial defects and the
lattice ($\gamma=0$, $\gamma'=0$), the various correlation functions
are simply given by the sum of the contributions from a lattice with
an equilibrium concentration of unbound dislocations (corresponding to
the {\em constrained} hexatic line liquid discussed by Marchetti and
Nelson
\cite{mcmdrn}) and from a liquid of defect lines.  The corresponding
elastic constants add in parallel, with
\begin{eqnarray}
& & \Big({1\over c_{11}^R(q_\perp)}\Big)_{\gamma=\gamma'=0}=
  \chi+{1\over c_{11}^{\rm MN}},\\  
& & \Big({1\over c_{44}^R(q_\perp)}\Big)_{\gamma=\gamma'=0}=
  {1\over K}+{1\over c_{44}^{\rm MN}},\\  
& & \Big({1\over c_{66}^R(q_\perp)}\Big)_{\gamma=\gamma'=0}=
  {1\over c_{66}^{\rm MN}},
\end{eqnarray}
where we have denoted by the superscript ``MN'' the elastic constants
of a lattice with an equilibrium concentration of unbound
dislocations, but with a constraint forbidding vacancy and
interstitial defects,\cite{mcmdrn} given by
\begin{eqnarray}
& &{1\over c_{11}^{\rm MN}}={1\over c_{11}}\bigg[1+{2 c_{66}^2\over 
    2c_{66}(c_{11}-c_{66})+c_{11}E_eq_\perp^2}\bigg]\,\label{c11hex}\\
& &{1\over c_{66}^{\rm MN}}=
   {1\over c_{66}}+{1\over E_e q_\perp^2}
 +{(c_{11}-2c_{66})\over 2c_{66}(c_{11}-c_{66})+c_{11}E_eq_\perp^2}\,\label{c66hex}\\
& &{1\over c_{44}^{\rm MN}}={1\over c_{44}}.\label{c44hex}
\end{eqnarray}
The elastic constants of a supersolid, i.e., the lattice with only
vacancy and interstitial defects, can be obtained by letting all the
dislocation core energies go to infinity
($E_e,E'_e,E_s,E'_s\rightarrow\infty$). It is easy to see that in this
limit we recover the results discussed in Sec. II.  Finally, when
$\chi^{-1}\rightarrow\infty$ and $K\rightarrow\infty$ vacancy and
interstitial defects are forbidden and the elastic constants reduce
again to those given in Eqs. \ref{c11hex}-\ref{c44hex}.

We recall that the coupling of dislocations to vacancy and
interstitial lines allows dislocations to relax stresses by climbing
out of the $({\bf b},\hat{\bf z})$ plane via the emission or
absorption of vacancies and interstitials {--} a mechanism for
relaxing stresses that is forbidden in the hexatic liquid of Marchetti
and Nelson
\cite{mcmdrn}. 
\begin{figure}[bth]
{\centering
\setlength{\unitlength}{1mm}
\begin{picture}(150,80)(0,0)
\put(-30,-70){\begin{picture}(150,80)(0,0) 
\includegraphics{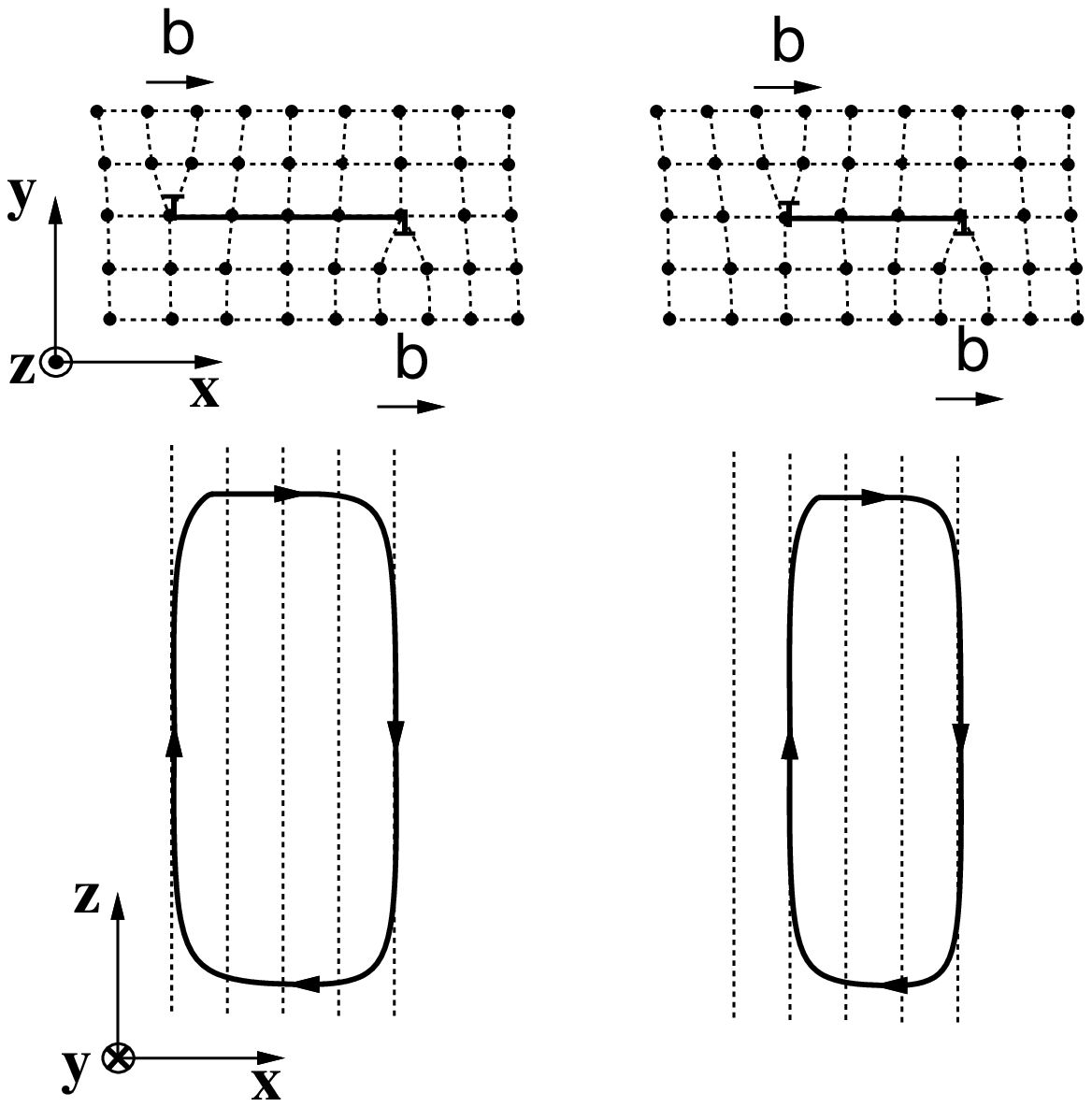}
\end{picture}}
\end{picture}}
Fig.5.{{\em Gliding} of an edge dislocation loop, allowed even in the
absence of vacancies and interstitials. For simplicity we have
sketched the dislocation loop for the case of a square lattice. The
dotted vertical lines refer to the defect-free square lattice and
serve as a guide for the eye.}
\end{figure}
Figure 5 shows a planar dislocation loop with ${\bf b}=b\hat{\bf x}$
of the type considered by Marchetti and Nelson \cite{mcmdrn}. 
This loop lies in its glide plane (the $xz$ plane) and can easily
relax a shear by gliding in this plane.  
The climbing of the same
dislocation loop out of its glide plane is described pictorially in
Fig.6.
\begin{figure}[bth]
{\centering
\setlength{\unitlength}{1mm}
\begin{picture}(150,70)(0,0)
\put(-28,-50){\begin{picture}(150,70)(0,0) 
\includegraphics{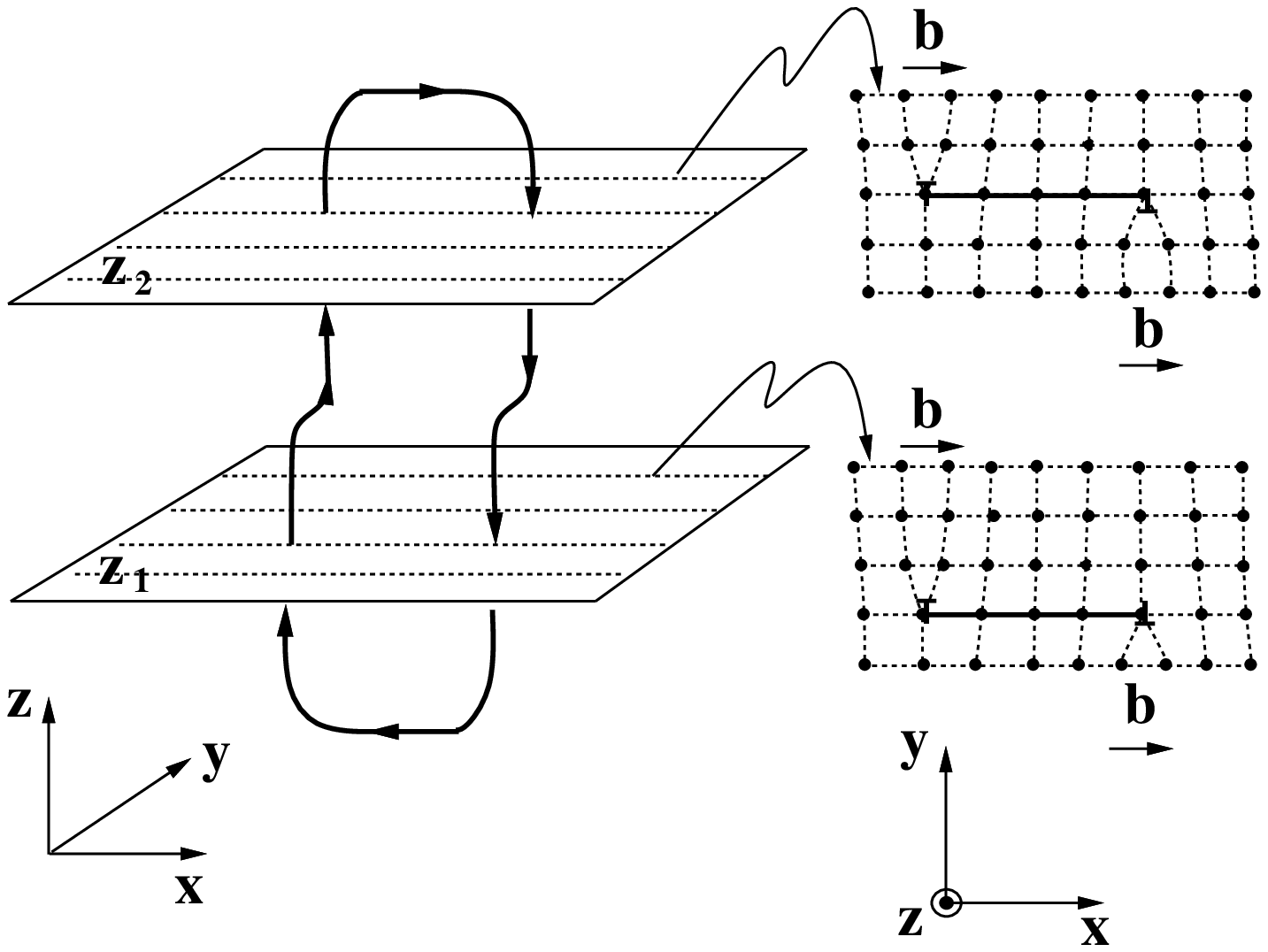}
\end{picture}}
\end{picture}}
Fig.6.{{\em Climbing} of an edge dislocation loop in a square lattice,
only allowed in the presence of vacancies and interstitials. The
Burger's vector of the loop is in the $+x$ direction. The lattice
configuration at two cross-sections $z_1$ and $z_2$ shows the climbing
in the $y$ direction, normal to the Burger's vector.}
\end{figure}

It is clear that climb can only occur via the emission or absorption
of vacancy or interstitial defects, as summarized by Eq.\ref{contdv}.
Such climb ``motion'' can occur in response to the force on the
dislocation loop resulting from applying a uniform external tilt to
the lattice, corresponding to a constant stress $\sigma^{\rm
ext}_{i\alpha}=\delta_{ix}\delta_{\alpha z}\sigma^{\rm ext}_{zx}$.
The force on a dislocation line due to a uniform stress
$\sigma^{\rm ext}_{i\alpha}$ is the familiar Peach-K\"ohler force, given by
\begin{equation}
\label{PKforce}
F^{\rm PK}_{\beta}=\epsilon_{\alpha\beta\gamma} \tau_\gamma\sigma^{\rm ext}_{i\alpha}b_i.
\end{equation}
Equation \ref{PKforce} differs slightly from the corresponding
expression found in textbooks as the stress tensor in a flux-line
lattice is not a symmetric (or even a square) matrix.  The
Peach-K\"ohler force on the rectangular loop shown in Fig.5 due to a
uniform tilt in the xy plane is normal to the plane of the loop (along
$y$) and there is no force on any segment of the loop parallel to $z$.
Specifically, for $\bbox{\tau}=\hat{\bf x}$, we find ${\bf F}^{\rm
PK,tilt}=-\sigma^{\rm ext}_{xz}b\hat{\bf y}$ and for
$\bbox{\tau}=-\hat{\bf x}$, we find ${\bf F}^{\rm PK,tilt}=\sigma^{\rm
ext}_{xz}b\hat{\bf y}$.  In other words, the Peach-K\"ohler force acts
as a couple and tries to rotate the loop out of its glide plane. The
screw components of the loop running along $\pm\hat{\bf x}$ can glide
in any plane that contains them. In particular, they will glide in the
$\hat{\bf y}$ or $-\hat{\bf y}$ direction under the action of the
force.  For this to happen the edge sections must climb out of the
glide plane by emitting or absorbing vacancies and interstitials, as
shown in Fig.6. By gliding out of the sample along the $y$ direction,
the screw dislocations can relax a uniform tilt of the lines towards
$\pm x$.  We note that if one thinks of $z$ as a fictitious time, the
constraint \ref{contdv} is formally identical to the temporal
continuity equation for the density of point vacancy/interstitial
defects, diffusing in the presence of dislocations in a
two-dimensional lattice
\cite{ZHN}.  Dislocations provide a source of point defects as they
climb across the sample.  The renormalization of the tilt modulus can
therefore occur only when vacancy and interstitial defects are
allowed.

It is also instructive to consider the behavior of another typical
dislocation loop, shown in Fig.7.  This loop has ${\bf b}=b\hat{\bf
x}$ and lies in the $yz$ plane.  The two edge segments running
parallel to $z$ lie in two different and parallel glide planes.  The
loop is closed by segments running along $y$ which are also edge
dislocations in nature, as ${\bf b}\perp\bbox{\tau}$. 

This is an example of a dislocation loop characterized by
$\epsilon_{ij}\alpha^\perp_{ij}\not=0$. Such a loop is not allowed in
the absence of vacancy and interstitials, as without such defects the
edge segments running along $y$ would require a row of vortex lines to
merge or split into single lines carrying twice or half a flux quantum
{--} an energetically forbidding configuration.  Denoting by $z_1<z_2$
the vertical location of the two segments of the loop running along
$y$, we can understand the existence of the loop as arising from a set
of interstitial defects that are randomly distributed in the $xy$
plane for $z<z_1$ and $z>z_2$, but organize themselves into a vortex
sheet for $z_1<z<z_2$, acting like an extra row of lines in this
region. In the region near $z_1$ and $z_2$, clearly $\partial_z
n_d\not=0$, corresponding to a nonvanishing value of
$\epsilon_{ij}\alpha^\perp_{ij}$.  Under a constant stress
$\sigma^{\rm ext}_{xz}$ from a uniform tilt applied to the system, the
Peach-K\"ohler force on the loop consists again of two forces of equal
magnitude applied on the sections running along $y$ and directed along
$\pm x$, as there is no force on the sections of the loop running
along $z$. Under the action of this couple, the loop rotates out of
its plane.  While this motion occurs in the glide plane for each
section of the loop, it requires ``diffusion'' of vacancies and
interstitials according to Eq. \ref{contdv}.
\begin{figure}[bth]
{\centering
\setlength{\unitlength}{1mm}
\begin{picture}(150,110)(0,0)
\put(-30,-70){\begin{picture}(150,110)(0,0) 
\includegraphics{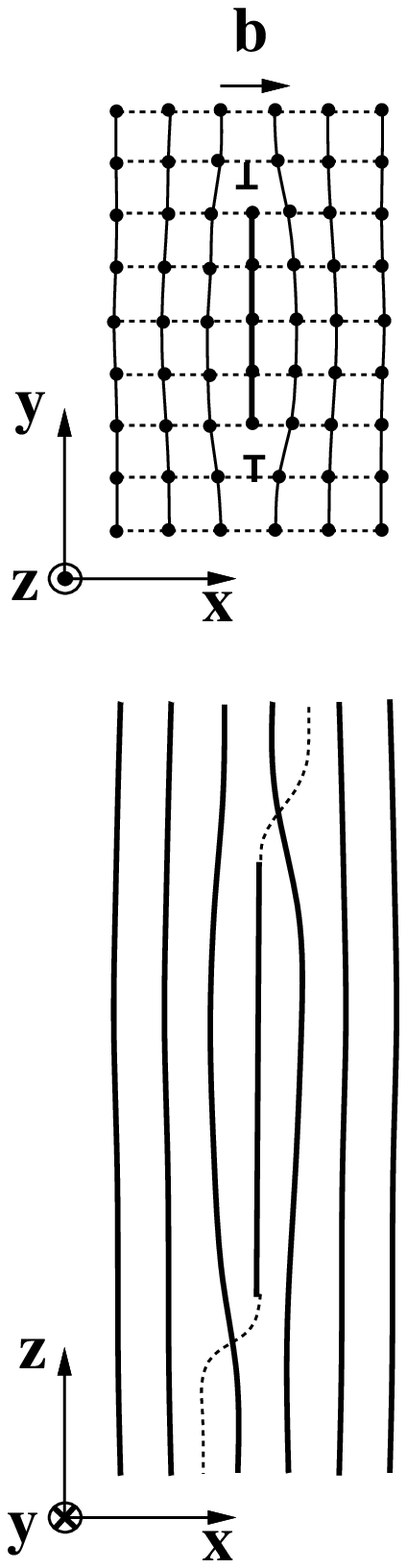}
\end{picture}}
\end{picture}}
Fig.7.{An edge dislocation loop described by a nonvanishing
$\alpha_{ij}^\perp\epsilon_{ij}$, allowed {\em only} in the presence
of vacancies and interstitials. The bottom figure showing a side view
of the loop emphasizes the equivalence between vacancy/interstitial
defects and a nonvanishing $\alpha_{ij}^\perp\epsilon_{ij}$. The loop
can be thought of as arising from an insertion of a row of
``finite-length vortices'', which correspond to a wandering of
interstitial defects.}
\end{figure}
The addition of dislocations also removes the nonanalyticity of the
transverse part of the tilt-tilt correlation function, present in both
the defect-free vortex lattice and the vortex supersolid.  In fact we
obtain
\end{multicols}
\begin{equation}
T_T(q_\perp, q_z=0)=n_0^2k_BT  
  {K+c_{44} + 2(E_s+E'_s)q^2_\perp -2\gamma'\over
     K c_{44} + 2K(E_s + E'_s)q_\perp^2-\gamma'^2 }.
\label{Ttqpqz=0}
\end{equation}
\begin{multicols}{2}
and
\begin{eqnarray}
\langle W^2\rangle&=&\lim_{q_\perp\rightarrow 0}T_T(q_\perp, q_z=0)\\\nonumber
  &=&{n_0^2k_BT  \over c_{44}^R},
\end{eqnarray}
with $c_{44}^R$ given by Eq. \ref{c44result}.  As indicated in
Eq. \ref{winding}, $\lim_{q_\perp\rightarrow 0}T_T(q_\perp, q_z=0)$
determines the winding number $\langle W^2\rangle$ and the
corresponding superfluid density $n_s$ of the equivalent boson system,
$\langle W^2\rangle=\hbar n_s/m$.  The analyticity of the transverse
tilt-tilt correlator at long wavelength indicates that $n_s=n_0$,
i.e., the bosons are in the superfluid state.  Conversely, this
corresponds to an entangled flux-line array, with no longitudinal
superconductivity.

For comparison, in the model of a {\em constrained} hexatic line
liquid, studied by Marchetti and Nelson,\cite{mcmdrn} where the
absence of vacancy/interstitial defects prevents dislocation loops
from climbing out of the $({\bf b},\hat{\bf z})$ plane, one obtains
\begin{equation}
\label{tilthex}
T_T^{\rm MN}(q_\perp, q_z=0)={n_0^2k_BT\over c_{44}+ 2(E_s + E'_s)q_\perp^2},
\end{equation}
and
\begin{equation}
\label{tilthex0}
\lim_{q_\perp\rightarrow 0}T_T^{\rm MN}(q_\perp, q_z=0)
={n_0^2k_BT\over c_{44}}.
\end{equation}
We find that independent of whether or not vacancy and interstitial
defects are included in the description of the hexatic vortex liquid,
the transverse tilt-tilt correlator is {\em analytic} and
$\lim_{q_\perp\rightarrow 0}T_T(q_\perp, q_z=0)$ is finite, indicating
that both systems are entangled. Vacancies and interstitials do,
however, decrease the tilt modulus, further enhancing the entanglement
of the vortex array.  Although vacancy and interstitial defects must
be incorporated in a consistent description of a flux-line hexatic,
they do not change the qualitative properties of the hexatic liquid,
which even in their presence maintains a vanishing long-wavelength
shear modulus and does not exhibit longitudinal superconductivity.

To further characterize orientationally ordered liquid it is useful to
define a characteristic length scale $\xi_{\perp}$, that determines
the typical transverse size of a disentangled flux-line bundle. This
is a region of the flux array where the lines remain locally
disentangled in the limit of infinite sample thickness along the field
($z$) direction.  In the absence of vacancies and interstitials,
Eq. (\ref{tilthex}) gives $\xi^{\rm MN}_\perp\sim
\sqrt{2(E_s+E'_s)/c_{44}}$. Using $E_s\sim E'_s\sim\sqrt
{c_{44}c_{66}}b^2$, we find $\xi^{\rm MN}_{\perp}\sim
a_0(c_{66}/c_{44})^{1/4}$, with $a_0=\sqrt{1/n_0}$ the mean
intervortex separation.  In flux-line arrays we typically have
$c_{66}<<c_{44}$ and the flux-line hexatic is entangled over all
macroscopic scales.

The coupling of dislocations to vacancy and interstitial defects
renormalizes this entangling length $\xi_{\perp}$, as can be seen from
Eq.\ref{Ttqpqz=0}, which for small $q_\perp$ is given by
\begin{equation} 
\label{tiltexpand}
T_T(q_\perp,0)\approx{n_0^2k_BT\over c_{44}^R}
\Big[1-q_\perp^2\xi_\perp^2\Big],
\end{equation}
with
\begin{equation}
\xi_{\perp}= \sqrt{c_{44}^R\over 2(E_s+E'_s)}{|K-\gamma'|\over
  |c_{44}+K-2\gamma'|}.
\end{equation}
It is easy to see that for all physical parameter values,
$\xi_\perp<\xi_\perp^{\rm MN}$. Not surprisingly, we find that
vacancies and interstitials therefore decrease the typical size of a
disentangled flux bundle.

Finally, we have also studied the effects of vacancy and interstitial
defects on the properties of the orientationally ordered hexatic
liquid.  Although the finite wavevector behavior is considerably
modified, we find that at long wavelength the effective hexatic
stiffness is still given by the expression obtained in
Ref. \onlinecite{mcmdrn}. This result is consistent with the lack of
long-scale orientational order and therefore a vanishing orientational
stiffness in the isotropic line-liquid of vacancies and interstitials.

\section{Conclusions}
In this paper we have studied the effects of vacancy and interstitial
line defects in the supersolid and the liquid phases of flux-line
arrays in the mixed state of type-II superconductors.

The transition from a fully ordered Abrikosov crystal phase into the
vortex supersolid phase takes place at a critical temperature or
magnetic flux density, at which vacancy and interstitial defects
proliferate, providing a mechanism for vortex entanglement and
yielding finite resistivity.  In order to study the long wavelength
elastic properties of the supersolid phase, we modeled the supersolid
as a lattice with an equilibrium concentration of unbound vacancy and
interstitial defect lines and computed the renormalization of the
elastic constants due fluctuations in the density and orientations of
these defects. As expected, at long wavelengths, we find that vacancy
and interstitial density fluctuations do {\em not} renormalize the
shear modulus, confirming the finiteness of the supersolid shear
modulus, which distinguishes it from the vortex liquid. In contrast,
these defect fluctuations yield a finite downward renormalization of
both the compressional and tilt moduli. The renormalization of the
tilt modulus stems from the fact that the liquid of vacancy and
interstitial defects promotes flux-line wandering. We explicitly
demonstrate that this defect-based vortex line delocalization
mechanism spoils the Meissner response to a shear tilt perturbation,
characteristic of the defect-free vortex lattice. In other words, the
vortex supersolid does not exhibit longitudinal superconductivity and,
in this respect, similarly to the vortex liquid, is always entangled.

It is clear that the vortex liquid phase, where unbound dislocation
loops have proliferated, is also characterized by a finite density of
vacancy and interstitials, which {\em must}, therefore, be included
for a consistent description of such a resistive phase. A model of the
vortex-line liquid as a solid with an equilibrium concentration of
unbound dislocation loops, but no vacancies and interstitials, was
previously studied by Marchetti and Nelson
\cite{mcmdrn}. As a consequence of no-vacancy/interstitial constraint
implicit in their model, only planar loops, that cannot climb out of
their glide plane, were included.  We have generalized this model by
allowing for a gas of vacancy and interstitial line defects, coupled
to the dislocation loop gas and the elastic (phonon) degrees of
freedom, to be present in the orientationally ordered vortex liquid.
We explicitly demonstrated that a finite concentration of these
vacancy and interstitial defects allows the dislocation loops to climb
out of their glide plane by emitting and absorbing vacancies and
interstitials, and allows for new types of edge dislocations
(otherwise forbidden), thereby significantly increasing the entropy
of topological defects characterizing the vortex liquid. This physical
result is mathematically summarized by Eq.\ref{contdv}, in which
non-planar dislocation loops and horizontal edge dislocations, with
Burger's vector in the xy-plane, act as sources for vacancy and
interstitial line defects. This resulting climb-like motion provides a
mechanism for the relaxation of an externally applied tilt, not
present in the model of Ref.\onlinecite{mcmdrn}.  We find, however,
that at long wavelengths there is {\em no independent} renormalization
of the tilt and compressional moduli by dislocations, other than that
already induced in the supersolid by fluctuations in the density and
orientation of vacancy and interstitial defects. This result is
somewhat puzzling and may be a consequence of the Debye-Huckel
approximation used to here.  We are currently investigating the
extension of the model presented here beyond the (quadratic)
Debye-Huckel treatment of defects in order to answer this question.
The coupling of dislocations to vacancy and interstitial defects does,
however, affect the response to a shear tilt perturbation that probes
longitudinal superconductivity. While both, the vortex supersolid and
the vortex liquid fail to exhibit longitudinal superconductivity, and
are therefore always entangled, in the supersolid it is only the
response of the vacancy and interstitial defects (which form only a
small fraction of the total vortex-flux density) that spoils the
Meissner effect, as
\begin{equation}
\lim_{q_\perp\rightarrow 0}\chi_T^{ss}(q_\perp,q_z=0)
=-{1\over 4\pi}\Big(1-{B^2\over 4\pi K}\Big),
\end{equation}
where $K$ is the tilt modulus of the vacancy-interstitial defect gas.
The transverse susceptibility $\chi^{ss}_T$, while larger than
$-1/4\pi${--} the value required for a perfect Meissner response {--}
remains negative (diamagnetic) and finite.  We expect therefore still
an appreciable screening of longitudinal currents, running along the
applied field direction.  In contrast, for the orientationally ordered
hexatic vortex liquid (where dislocations loops also unbind), we find
\begin{equation}
\lim_{q_\perp\rightarrow 0}\chi_T^{hex}(q_\perp,q_z=0)
=-{1\over 4\pi}\Big(1-{B^2\over 4\pi c_{44}^R}\Big),
\end{equation}
where $c_{44}^R$ is the significantly reduced vortex liquid tilt
modulus, given by Eq.\ref{c44result}.  At all, but very low vortex
densities $c_{44}^R\approx B^2/4\pi$ and the diamagnetic
susceptibility is therefore vanishingly small. Hence, as expected, the
fully entangled vortex liquid has essentially normal response to a
shear tilt perturbation and exhibits no longitudinal
superconductivity.

We hope that the results for the properties of the supersolid and
hexatic phases, and as well as for the transitions scenarios from and
into them, presented here, will serve as a useful guide for
identifying these exotic phases in experiments\cite{forgan} and
simulations\cite{chenteitel} of vortex systems. It is also likely that
the equilibrium analytical description developed here will be useful
for incorporating defects into the current elastic theories of {\em
driven} vortex lattices\cite{SV,BMR}, as well as for interpreting and
characterizing the results of experiments\cite{andrei} and numerical
simulations\cite{moon,nori} on these rich nonequilibrium systems.

\section{Acknowledgments}
We both thank The Institute for Theoretical Physics at the University
of California Santa Barbara, and the organizers of the Vortex Workshop
held there, where this work was initiated, for their hospitality and
financial support under NSF Grant No. PHY94-07194. We have benefited
from conversations with David Nelson, Daniel Fisher, Erwin Frey and
John Toner.  MCM was supported by the National Science Foundation at
Syracuse through Grants DMR-9730678 and DMR-9805818. LR was
financially supported by the National Science Foundation CAREER award,
through Grant \# DMR-9625111, and by the A.P. Sloan and the David and
Lucile Packard Foundations.

\section{Appendix A: Details of coupling constants appearing in 
the supersolid model} 

In this appendix we study the phenomenological couplings that appear
in our model of a supersolid, defined by the effective free energy
$F_{ss}$, Eq.\ref{Fss}. Our goal is to determine the range of values that these
parameters can assume in a physically realistic model of a
supersolid.
In Sec.\ref{CRsupersolid_functions} we found that the
effective supersolid moduli $c_{11}^R$ and $c_{44}^R$,
Eqs.\ref{c11ren},\ref{c44ren} vanish at intermediate values of the
couplings $\gamma$ and $\gamma'$, suggesting a vortex density
instability within the supersolid phase. The purpose of this appendix
is to show that such an instability is only apparent and arises 
from our definition of parameters.

As we have seen in Sec.\ref{model_supersolid}, the supersolid
effective free energy consists of the elastic lattice part, $F_{\rm
latt}$, the vacancy-interstitial defect part, $F_{\rm d}$, and a
coupling between the elastic and defect degrees of freedom, 
$F_{\rm int}$, given in Eq.\ref{freeint}. The form of $F_{\rm int}$ was
dictated by symmetry, with $\gamma$ and
$\gamma'$ unknown phenomenological parameters. We argue
here, however, that in a more realistic model of a supersolid, the
coupling between elastic degrees of freedom and defects arises from fluctuations 
in the magnetic
induction, $\delta{\bf B}$, and therefore should be written as an
expansion in powers of $\delta{\bf B}$. To lowest (quadratic) order,
such energetic contribution, when expressed in terms of vortex line
density fluctuations $\delta n$ and ${\bf t}$, lead to $\tilde{F}_{\rm
int}$ given by
\begin{equation}
\label{tildeFint}
\tilde{F}_{\rm int}={1\over 2 n_0^2}\int d{\bf r}\Big[\alpha(\delta
n)^2 + \alpha'({\bf t})^2\Big]\;.
\end{equation}
Inserting the expressions for $\delta n$ and ${\bf t}$ given in 
Eqs.\ref{deltan} and \ref{deltat} in $\tilde{F}_{\rm int}$, expanding
the resulting expression and combining it with $F_{\rm latt}$ and
$F_{\rm d}$, we find $\tilde{F}_{\rm ss}=F_{\rm latt}+F_{\rm
d}+\tilde{F}_{\rm int}$
\end{multicols}
\begin{equation}
\label{tildeFss}
\tilde{F}_{\rm ss}={1\over 2}\int d{\bf r}\Big
[2\tilde{c}_{66}u_{ij}^2+(\tilde{c}_{11}-2\tilde{c}_{66})u_{kk}^2
+\tilde{c}_{44}(\partial_z{\bf u})^2+{1\over\tilde{\chi} n_0^2}(\delta
n_d)^2 + {\tilde{K}\over n_0^2}({\bf t}_d)^2
+{2\tilde{\gamma}\over n_0}\delta n_d\bbox{\nabla}\cdot{\bf u}
+{2\tilde{\gamma}'\over n_0}{\bf t}_d\cdot\partial_z{\bf u}\Big]\;.
\end{equation}
\begin{multicols}{2}
\noindent
Clearly Eq. \ref{tildeFss} has the same functional form as $F_{\rm ss}$ studied
in the main text. The coupling constants, however, are given by 
\begin{mathletters}
\begin{eqnarray}
\tilde{c}_{11}&=&c_{11}+\alpha\;,\\
\tilde{c}_{66}&=&c_{66}\;,\\
\tilde{c}_{44}&=&c_{44}+\alpha'\;,\\
\tilde{\chi}^{-1}&=&\chi^{-1}+\alpha\;,\\
\tilde{K}&=&K+\alpha'\;,\\
\tilde{\gamma}&=&-\alpha\;,\\
\tilde{\gamma}'&=&\alpha'\;.
\end{eqnarray}
\label{tilde_parameters}
\end{mathletters}
and they all depend on the two independent parameters $\alpha$ and
$\alpha'$, which are always positive.

We can now reexpress the renormalized supersolid moduli $c_{11}^R$ and
$c_{44}^R$, given by Eqs.\ref{c11ren} and \ref{c44ren}, respectively,
in terms of the couplings of the model defined here, with the result
\begin{mathletters}
\begin{eqnarray}
c_{11}^R&=&{c_{11}\chi^{-1}+(c_{11}+\chi^{-1})\alpha\over
c_{11}+\chi^{-1}}\;,\\ 
c_{44}^R&=&{c_{44}K+(c_{44}+K)\alpha'\over
c_{44}+K}\;,
\end{eqnarray}
\label{c_alpha}
\end{mathletters}
which clearly do not vanish (or diverge) for any positive values of
$\alpha$ and $\alpha'$. Hence the instability found in
Sec.\ref{CRsupersolid_functions} was spurious, an artifact of
expressing our results in terms of $\gamma$ and $\gamma'$ and allowing
these coupling constants to  access values that are unphysical in a generic
model of a supersolid.
\end{multicols}

\section{Appendix B: The defect free energy}

Here we give the expressions for the various kernels contained in the
defect free energy of Eq.\ref{freealpha}. The nonlocal kernel
$\tilde{R}_{\alpha i,\beta j}$ is given by
\begin{equation}
\label{Raibj}
\tilde{R}_{\alpha i,\beta j}({\bf q})=
{1\over  q^2}B_{\alpha i,\beta j}({\bf q})
    +\Delta_{\alpha i,\beta j}({\bf q})+E_{\alpha i,\beta j},
\end{equation}
where 
\begin{equation}
\label{Baibj}
B_{\alpha i,\beta j}({\bf q})=\Big[C_{\alpha i\beta j}
  -C_{\alpha i\gamma k}q_\gamma(A^{-1})_{kl} q_\eta C_{\eta l\beta j}\Big]
   \epsilon_{\alpha\lambda\mu}\hat{q}_\lambda
           \epsilon_{\beta\xi\nu}\hat{q}_\xi,
\end{equation}
describes the long-range interaction between dislocation loops in the
absence of vacancy and interstitial defects, while
\begin{eqnarray}
\label{Deltaaibj}
\Delta_{\alpha i,\nu j}({\bf q})=& &{1\over q_\perp^2}\Big(K-{\gamma'^2q_z^2\over\Gamma_L}\Big)
   \epsilon_{z\alpha i}\epsilon_{z\beta j}
  -{\gamma'\hat{q}_{\perp k}\over q^2}\big[\epsilon_{zk\alpha}\hat{q}_{\perp i}\epsilon_{z\beta j}
             +\epsilon_{zk\beta}\hat{q}_{\perp j}\epsilon_{z\alpha i}\big]\\
 & & +{\gamma' q_z q_\mu q_\gamma\hat{q}_{\perp k}\over q^2q_\perp\Gamma_L}
      \big[C_{\mu k\nu i}\epsilon_{\nu\gamma\alpha}\epsilon_{z\beta j}
           +C_{\mu k\nu j}\epsilon_{\nu\gamma\beta}\epsilon_{z\alpha i}\big].
\end{eqnarray}
is the part of such interaction mediated by vacancy and interstitial
defects.  The matrix $E_{\alpha i,\beta j}$ describes the dislocation
core energy and is given by
\begin{equation}
E_{\alpha i,\beta j}=2E_e\delta_{\alpha z}\delta_{\beta z}\delta_{ij}
   + 2E_s\delta_{\alpha i}\delta_{\beta j}
   +2E'_s\delta_{\alpha\beta}(1-\delta_{\alpha z}\delta_{\beta z})\delta_{ij}
   +2E'_e\epsilon_{z\alpha i}\epsilon_{z\beta j}.
\end{equation}
The effective interaction between the vacancy and interstitial defects
is described by the two scalar kernels
\begin{equation}
\label{Aq}
A({\bf q})={1\over n_0^2}\Big[\chi^{-1}+K{q_z^2\over q_\perp^2}
  -\big(\gamma-\gamma'q_z^2/q_\perp^2\big)^2{q_\perp^2\over\Gamma_L}\Big],
\end{equation}
and
\begin{equation}
\label{Cq}
C({\bf q})={1\over n_0^2}\Big[K-{{\gamma'}^2q_z^2\over\Gamma_T}\Big].
\end{equation}
Finally, the tensors $D_{\alpha i}$ and $G_{\alpha i}$ describe the
coupling between the dislocation loop gas and the liquid of
vacancy-interstitial defect lines, and are given by
\begin{eqnarray}
& &D_{\alpha i}({\bf q})={1\over n_0 q}\Big\{\big[\gamma\delta_{\beta i}-\gamma'\delta_{\beta z}
  \hat{q}_{\perp i}q_z/q_\perp\big]
 -{ q_\perp\over\Gamma_L}\big(\gamma -\gamma'q_z^2/q_\perp^2\big)q_\mu\hat{q}_{\perp i}
  C_{\mu j\beta i}\Big\}\epsilon_{\beta\gamma\alpha}\hat{q}_\gamma
  -{q_z\over n_0}\Big[{K\over q_\perp^2}+{\gamma'\over\Gamma_L}\big(\gamma -\gamma'q_z^2/q_\perp^2\big)\Big]
              \epsilon_{z\alpha i},\\
& &G_{\alpha i}({\bf q})={\gamma'\over n_0q}\Big[\gamma\delta_{\beta z}\hat{C}_i
    -{q_z\over\Gamma_T}q_\mu\hat{C}_j C_{\mu j\beta i}\Big]
     \epsilon_{\beta\gamma\alpha}\hat{q}_\gamma,
\end{eqnarray}
with $\hat{C}_i=\epsilon_{ij}\hat{q}_{\perp j}$ and 
$\hat{q}_\xi=q_\xi/q$,

\begin{multicols}{2}


\end{multicols} 

\begin{references}

\bibitem{huse_radzihovsky}
D. Huse and L. Radzihovsky, in Proceedings
of 1993 Altenberg Summer School, {\em Fundamental Problems in
Statistical Mechanics VIII}, edited by H. van Beijeren and M. H. Ernst
(Elsevier, Netherlands).

\bibitem{blatter}
G. Blatter, M.V. Feigel'man, V.B. Geshkenbein,
A.I. Larkin, and V.M. Vinokur, Rev. Mod. Phys. {\bf 66}, 1125 (1994).

\bibitem{brandt_review}
E. H. Brandt, Rep. Prog. Phys. {\bf 58}, 1465 (1995).

\bibitem{nelson_seung}
D.R. Nelson, Phys. Rev. Lett. {\bf 60}, 1973
(1988); D.R. Nelson and S. Seung, Phys. Rev. B {\bf 39}, 9153 (1989).

\bibitem{Hc1comment} 
As first pointed out in Ref.\onlinecite{nelson_seung}, one also
expects a vortex liquid state to exist just above the $H_{c1}(T)$
curve, in very clean superconductor samples.

\bibitem{FFH} 
M. P. A. Fisher, Phys. Rev. Lett. {\bf 62}, 1415 (1989);
D. S. Fisher, M. P. A. Fisher, and D. A. Huse, Phys. Rev. B {\bf 43}, 
130 (1991).

\bibitem{Larkin} 
A. Larkin, Sov. Phys. JETP {\bf 31}, 784 (1970);
A.I. Larkin and Y.N. Ovchinnikov, Sov. Phys. JETP {\bf 38}, 854
(1974).


\bibitem{BraggGlass}
T. Giamarchi and Le Doussal, Phys. Rev. Lett. {\bf 72}, 1530 (1994);
M.J.P. Gingras and D.A. Huse, Phys. Rev. B {\bf 53}, 15183;
D. S. Fisher, Phys. Rev. Lett. {\bf 78}, 1964 (1997). In ultra-clean
superconductors, the distinction between the conventional ordered
Abrikosov lattice and the Bragg glass appears, however, only beyond a
very large Larkin scale.\cite{Larkin} Analogous topologically ordered,
but elastically disordered phases also appear in other randomly pinned
periodic systems, such as for example smectic liquid crystals confined
in aerogel: L. Radzihovsky and J. Toner, Phys. Rev. Lett. {\bf 79},
4214 (1997); B. Jacobsen, K. Saunders, L. Radzihovsky and J. Toner,
unpublished.

\bibitem{Gammel} 
P.L. Gammel, L.F. Scheemeyer, J.V. Waszczak, and D.J. Bishop, Phys. Rev. Lett.
{\bf 61}, 1666 (1988).

\bibitem{Koch} 
R.H. Koch et al., Phys. Rev. Lett. {\bf 63}, 1511 (1989).

\bibitem{Zeldov} E. Zeldov et. al, Nature {\bf 382}, 791 (1996).

\bibitem{melting} Although a clean vortex crystal appears to melt via a
first order transition, as first observed in Ref.\onlinecite{Zeldov},
a fluctuation-driven {\em continuous} melting transition is
theoretically possible, as demonstrated for model systems in
Ref.\onlinecite{continuous_melting}.

\bibitem{continuous_melting}  L. Radzihovsky, 
Phys. Rev. Lett. {\bf 74}, 4722 (1995); L. Balents and L. Radzihovsky,
Phys. Rev. Lett. {\bf 76}, 3416 (1996). S.A. Ktitorov, B.N.Shalaev,
and L. Jastrabik, Phys. Rev. B {\bf 49}, 15248 (1994).

\bibitem{KT} 
J.M. Kosterlitz and D.J. Thouless, J. Phys. C{\bf 6}, 1181 (1973);
B. I. Halperin and D. R. Nelson, Phys. Rev. Lett. {\bf 41}, 121
(1978); D. R. Nelson and B. I. Halperin, Phys. Rev. B {\bf 19}, 2457
(1979); A. P. Young, Phys. Rev. B {\bf 19}, 1855 (1979).

\bibitem{dsfisher} 
D.S. Fisher, Phys. Rev. B {\bf 22}, 1190 (1980).

\bibitem{mcmdrn} M.C. Marchetti and D.R. Nelson, Phys. Rev. B {\bf
41}, 1910 (1990).

\bibitem{NelsonToner} 
D.R. Nelson and J. Toner, Phys. Rev. B {\bf 24}, 363 (1981).

\bibitem{martin} P. C. Martin, O. Parodi, and P.S. Pershan,
Phys. Rev. A {\bf 6}, 2401 (1972).

\bibitem{MPAFisherLee} 
M. P. A. Fisher and D. H. Lee, Phys. Rev. B {\bf 39},
2756 (1989).

\bibitem{fnf} 
E. Frey, D.R. Nelson and D.S. Fisher, Phys. Rev. B {\bf 49},
9723 (1994).

\bibitem{glazman}
L.I. Glazman and A.E. Koshelev, Phys. Rev. B {\bf 43}, 2835 (1991).

\bibitem{fb} E. Frey and L. Balents, Phys. Rev. B {\bf 55}, 1050 (1997).

\bibitem{fuchs}
D.T. Fuchs, E. Zeldov, T. Tamegai, S. Ooi, M. Rappaport, and
H. Shtrikman, Phys. Rev. Lett. {\bf 80}, 4971 (1998).

\bibitem{forgan}
E.M. Forgan {\it et al.}, Czech. J. Phys. {\bf 46}, 1571 (1996).

\bibitem{melting_comment} 
Since in type I melting two order parameters (vacancy-interstitial
density and dislocation density) with unrelated symmetry order
simultaneously, we expect type I melting to be 1st order.  In
contrast, type II melting can be continuous, consisting of $3$
intermediate transitions: solid-to-supersolid, supersolid-to-hexatic,
and hexatic-to-(isotropic) liquid (see Ref.\onlinecite{continuous_melting}).

\bibitem{defects_comment} In a vortex liquid dislocations are, by definition,
unbound. Furthermore, a pair of dislocations forms a vacancy or an
interstitial defect.  We expect therefore that there will always be a finite
density of vacancy and interstitial defects in the liquid phase.

\bibitem{notation_comment} The relation to the Lam\'e coefficients
$\mu$ and $\lambda$, commonly used in the study of elasticity of
solids is: $c_{66}=\mu$ and $c_{11}=2\mu+\lambda$.

\bibitem{brandt} E.H. Brandt and U. Essman, Phys. Status Solidi B {\bf
144}, 13 (1987).

\bibitem{dsfisher_elasticity} D.S. Fisher, in {\it Phenomenology and Applications of High-Temperature 
Superconductors}, K.S. Bedell et al., eds. (Addison-Wesley, 1992), p. 287.


\bibitem{ledou} 
D.R. Nelson and P. Le Doussal, Phys. Rev. B {\bf 42}, 10113 (1990).

\bibitem{foota} 
More generally the relationship between the
hydrodynamic fields and the local induction is nonlocal and can be
obtained by solving the anisotropic London equation, as shown for
instance in Ref.\onlinecite{ledou}.

\bibitem{chenteitel} 
T. Chen and S. Teitel, Phys. Rev. B {\bf 55}, 15197 (1997).

\bibitem{BoseGlass} D. R. Nelson and V. M. Vinokur, Phys. Rev. Lett.
{\bf 68}, 2398 (1992); Phys. Rev. B. {\bf 48}, 13060 (1993).

\bibitem{pollock} E.L. Pollock and D.M. Ceperley, Phys. Rev. B{\bf
36}, 8343 (1987).

\bibitem{pbmcm} P. Benetatos and M.C. Marchetti, to appear
in Phys. Rev. B (cond-mat/9808270).

\bibitem{RFhydro} L. Radzihovsky and E. Frey, 
Phys. Rev. B, {\bf 48}, 10357 (1993).

\bibitem{MNhydro} M.C. Marchetti and D.R. Nelson, Physica C {\bf 174},
40 (1991).

\bibitem{ClareYu} H.M. Carruzzo and C.C. Yu, Phil. Mag. B {\bf 77}, 1001 (1998).

\bibitem{ZHN} A. Zippelius, B.I. Halperin and D.R. Nelson,
Phys. Rev. B {\bf 22}, 2514 (1980).

\bibitem{comment_stability} We note in passing that from this
expression we see that the stability of a defect-free crystal is
determined by the two conditions: (i) $c_{66}>0$ and (ii)
$c_{11}>c_{66}$ (or equivalently in Landau and Lifshitz notation
$\mu>0$ and $\mu+\lambda>0$), derived from the requirement of a finite
zero mode response to a {\em uniform} stress. This condition is
different and more stringent than that imposed by the requirement of a
finite response of the finite wavevector bulk modes ($u_T({\bf q})$
and $u_L({\bf q})$), obtained from the propagator of $F_{el}$
Eq.\ref{Fss}, which gives: (i) $c_{66}>0$ and (ii) $c_{11}>0$.

\bibitem{nabarro}
F.R.N. Nabarro and A.T. Quintanilha, in {\it Dislocations in Solids},
edited by F.R.N. Nabarro (North-Holland, Amsterdam, 1980), Vol. 5.

\bibitem{brandtent}
E.H. Brandt, Phys. Rev. B {\bf 34}, 6514 (1986); Jpn. J. Appl. Phys.
{\bf 26}, 1515 (1987).

\bibitem{notetilde}
These renormalized moduli can also be obtained by simply
integrating out the {\em defect} degrees of freedom $\delta n_d$ and
$t_d^T$ inside the partition function, and identifying the
corresponding effective elastic moduli in the remaining effective
elastic free energy functional.

\bibitem{SV} S. Scheidl and V.M. Vinokur, unplublished (cond-mat/9702014).

\bibitem{BMR} L. Balents, M.C. Marchetti, and L. Radzihovsky, 
Phys. Rev. B {\bf 57}, 7705 (1998).

\bibitem{andrei} E.Y. Andrei, G. Deville, D.C. Glattli, F.I.B. Williams, 
E. Paris, and B. Etienne, Phys. Rev. Lett. {\bf 60}, 2765 (1988).

\bibitem{moon} K. Moon, R. T. Scalettar, and G. Zim\'anyi, Phys. Rev. Lett. 
{\bf 77}, 2778 (1996).

\bibitem{nori} C.J. Olson, C. Reichhardt, and F. Nori, Phys. Rev. Lett. 
{\bf 80}, 2197 (1998), ibid. {\bf 81}, 3757 (1998).

\end{references}
\end{document}